\documentclass[twocolumn]{aastex631}
\usepackage{amsmath}

\newcommand{\citepe}[1]{\citep[e.g.,][]{#1}}

\newcommand{\comment}[1]{}
\usepackage{bm}
\newcommand{\Add}[1]{\textcolor{black}{#1}}
\newcommand{\Erase}[1]{}

\newcommand{\M}{{\ $\mid$}}

\newcommand{\W}{{$\lambda$}}

\newcommand{\OII}{[O\,\textsc{ii}]}

\newcommand{\Hb}{H$\beta$}
\newcommand{\Hc}{H$\gamma$}

\newcommand{\OIII}{[O\,\textsc{iii}]}

\newcommand{\NeIII}{[Ne\,\textsc{iii}]}

\newcommand{\HI}{H\,\textsc{i}}

\newcommand{\q}{$q_{\rm ion}$}

\graphicspath{{./}{figures/}}
\shorttitle{
Measurements of $x_{\rm HI}$ and Ionized Bubble Sizes at $z=7-12$
}
\shortauthors{Umeda et al.}

\begin{document}
%\linenumbers

\title{
JWST Measurements of Neutral Hydrogen Fractions and Ionized Bubble Sizes at $z=7-12$\\
Obtained with Ly$\alpha$ Damping Wing Absorptions in 27 Bright Continuum Galaxies
}

\author[0009-0008-0167-5129]{Hiroya Umeda}
\affiliation{Institute for Cosmic Ray Research,
The University of Tokyo,
5-1-5 Kashiwanoha, Kashiwa,
Chiba 277-8582, Japan}
\affiliation{Department of Physics, Graduate School of Science, The University of Tokyo, 7-3-1 Hongo, Bunkyo, Tokyo 113-0033, Japan}
\email{ume@icrr.u-tokyo.ac.jp}

\author[0000-0002-1049-6658]{Masami Ouchi}
\affiliation{National Astronomical Observatory of Japan, 2-21-1 Osawa, Mitaka, Tokyo 181-8588, Japan}
\affiliation{Institute for Cosmic Ray Research,
The University of Tokyo,
5-1-5 Kashiwanoha, Kashiwa,
Chiba 277-8582, Japan}
\affiliation{Kavli Institute for the Physics and Mathematics of the Universe (WPI), 
University of Tokyo, Kashiwa, Chiba 277-8583, Japan}
\author[0000-0003-2965-5070]{Kimihiko Nakajima}
\affiliation{National Astronomical Observatory of Japan, 2-21-1 Osawa, Mitaka, Tokyo 181-8588, Japan}

\author[0000-0002-6047-430X]{Yuichi Harikane}
\affiliation{Institute for Cosmic Ray Research,
The University of Tokyo,
5-1-5 Kashiwanoha, Kashiwa,
Chiba 277-8582, Japan}
\affiliation{Department of Physics and Astronomy, University College London, Gower Street, London WC1E 6BT, UK}

\author[0000-0001-9011-7605]{Yoshiaki Ono}
\affiliation{Institute for Cosmic Ray Research,
The University of Tokyo,
5-1-5 Kashiwanoha, Kashiwa,
Chiba 277-8582, Japan}

\author[0000-0002-5768-8235]{Yi Xu}
\affiliation{Institute for Cosmic Ray Research, The University of Tokyo, 5-1-5 Kashiwanoha, Kashiwa, Chiba 277-8582, Japan}
\affiliation{Department of Astronomy, Graduate School of Science, The University of Tokyo, 7-3-1 Hongo, Bunkyo, Tokyo 113-0033, Japan}

\author[0000-0001-7730-8634]{Yuki Isobe}
\affiliation{Institute for Cosmic Ray Research,
The University of Tokyo,
5-1-5 Kashiwanoha, Kashiwa,
Chiba 277-8582, Japan}
\affiliation{Department of Physics, Graduate School of Science, The University of Tokyo, 7-3-1 Hongo, Bunkyo, Tokyo 113-0033, Japan}

\author[0000-0003-3817-8739]{Yechi Zhang}
\affiliation{Institute for Cosmic Ray Research,
The University of Tokyo,
5-1-5 Kashiwanoha, Kashiwa,
Chiba 277-8582, Japan}
\affiliation{Department of Astronomy, Graduate School of Science, The University of Tokyo, 7-3-1 Hongo, Bunkyo, Tokyo 113-0033, Japan}

\begin{abstract}
%250 words
We present volume-averaged neutral hydrogen fractions $x_{\rm \HI}$ and ionized bubble radii $R_{\rm b}$ measured with Ly$\alpha$ damping wing absorption of galaxies at the epoch of reionization. We combine JWST/NIRSpec spectra taken by CEERS, GO-1433, DDT-2750, and \Add{JADES} programs, and obtain \Add{a sample containing} 2\Add{7} bright UV-continuum \Add{($M_{\rm UV}<-18.5~{\rm mag}$)} galaxies at $7<z<12$. We construct 4 composite spectra binned by redshift, and find the clear evolution of \Add{softening break} towards high redshift at the rest-frame $1216$ {\AA}\Add{,} suggesting the increase of Ly$\alpha$ damping wing absorption. We estimate Ly$\alpha$ damping wing absorption in the \Add{galaxy} spectra with realistic templates including \Add{Ly$\alpha$ emission and }circum-galactic medium absorptions. Assuming the standard inside-out reionization picture \Add{having an ionized bubble with radius $R_b$ around a galaxy embedded in the intergalactic medium with $x_{\rm \HI}$}, we obtain $x_{\rm \HI}$ ($R_{\rm b}$) values \Add{generally} increasing (decreasing) from \Add{$x_{\rm \HI}={0.53}^{+0.18}_{-0.47}$ to ${0.92}^{+0.08}_{-0.10}$ ($\log R_{\rm b}={1.67}^{+0.14}_{-0.16}$ to ${-0.69}^{+0.89}_{-0.24}$ comoving Mpc) at redshift $7.12^{+0.06}_{-0.08}$ to \Add{$9.91^{+1.49}_{-1.15}$}}. The redshift evolution of $x_{\rm \HI}$ indicates \Add{a} moderately late reionization history consistent with the one \Add{previously} suggested from the electron scattering of cosmic microwave background and the evolution of UV luminosity function with an escape fraction $f_{\rm esc}\Add{\sim 0.2}$. Our ${R_{\rm b}}$ measurements \Add{suggest that bubble sizes} could be up to \Add{a few} dex larger than the cosmic average values estimated by analytic calculations for a given $x_{\rm \HI}$, while our $R_{\rm b}$ measurements are \Add{roughly} comparable with the values for merged ionized bubbles around bright galaxies predicted by recent numerical simulations. 
\end{abstract}

\section{Introduction} \label{intro}
Cosmic reionization scenarios have been investigated by estimating the neutral hydrogen fraction $x_{\rm \HI}$ of the universe at different redshifts. \Add{Most powerful way} to probe the  neutral hydrogen at the epoch of reionization is to look at Ly$\alpha$ damping wing absorption. Numerous works have investigated $x_{\rm \HI}$ from Ly$\alpha$ damping wing absorption using the brightest sources such as quasars (QSOs) and gamma-ray burst (GRBs) at $z\sim6-8$ \citep[e.g.,][]{M11,B18,2013MNRAS.428.3058S,2018ApJ...864..142D,2019MNRAS.484.5094G,2020ApJ...896...23W,2006PASJ...58..485T,2014PASJ...66...63T}. \Add{With the bright rest-frame UV-continuum, we can detect Ly$\alpha$ damping wing absorption by the intergalactic medium (IGM). $x_{\rm \HI}$ value at different redshift have been estimated by fitting the damping wing absorption profile for different $x_{\rm \HI}$ values.} However, the number density of QSOs and GRBs are expected to be very small at the higher redshift, inhibiting us from accessing the ionization state of IGM at $z\gtrsim7$ \citep[e.g.,][]{2019ApJ...884...30W,2023arXiv230511225M}.\par

To overcome the problem of lacking the bright objects at the high redshift, we investigate the fainter but much more \Add{numerous} galaxies as alternative target to infer $x_{\rm \HI}$ from Ly$\alpha$ damping wing absorptions. To directly observe the Ly$\alpha$ damping wing absorption in the spectrum of galaxies at $z>7$, we need to detect \Add{their} rest-frame UV continuum. Spectroscopically detecting the rest-frame UV continuum at such a high redshift is very challenging because it requires deep spectroscopic data in near infrared \Add{(NIR)}. However, recent observations by \Add{the} James Webb Space Telescope \Add{\citep[JWST;][]{2023PASP..135f8001G,2023PASP..135d8001R,2023PASP..135b8001R,2022A&A...661A..80J,2022A&A...661A..81F,2023PASP..135c8001B}} have shown that \Add{it is} capable of deep NIR spectroscopic observations. As presented in recent works, low-resolution (i.e., $R\sim100$) NIRSpec/PRISM spectroscopic data is especially suited for detecting faint rest-frame UV continuum \citep[e.g.,][]{CL23, Nk23,Ha23,AH23a,AH23b,H23,W23,RB23}.\par

\cite{CL23} are the first to demonstrate the power of the PRISM data to estimate Ly$\alpha$ damping wing absorption in \Add{the UV continua of high redshift galaxies}. \cite{CL23} \Add{have} demonstrated the potential of the NIRSpec PRISM data to estimate \Add{the value of} $x_{\rm \HI}$. However, their constraint on $x_{\rm \HI}$ is not very strict because the redshift for the object is not constrained due to \Add{the lack of emission line detections}. Recently, \cite{H23} have incorporated Ly$\alpha$ damping wing absorption in the spectral fitting of spectroscopically confirmed $z=10.17$ lensed galaxy and constrained $x_{\rm \HI}>0.9$ with a one sigma significance. In our work, we also take advantage of the \Add{multiple} PRISM spectroscopy data with the redshift confirmation obtained by multiple surveys to systematically infer the $x_{\rm \HI}$ values at various redshift above $z=7$ to capture the redshift evolution of the damping wing profile and corresponding $x_{\rm \HI}$ values.\par

This paper is constructed as follows. In Section \ref{data}, we describe our galaxy sample and the reduction of corresponding spectroscopic data. We construct composite spectra binned by redshift and fit for the Ly$\alpha$ damping wing absorption seen in the composite spectra in Section \ref{met}. We discuss the implications of the $x_{\rm \HI}$ estimates on understanding cosmic reionization history and future prospect in Section \ref{disc}. We summarize this paper in Section \ref{conc}. In this paper, we use the Planck cosmological parameter sets of the TT + TE + EE+ lowE + BAO + lensing result \citep{Planck}: $h=0.6766$, $\Omega_m=0.3103$, $\Omega_\Lambda=0.6897$, $\Omega_b h^2=0.02234$, and $Y_p=0.248$. All magnitudes are in the AB system \citep{1983ApJ...266..713O}.\par
\section{Data and Sample} \label{data}
\begin{figure}
\centering
\begin{center}
\includegraphics[width=\linewidth]{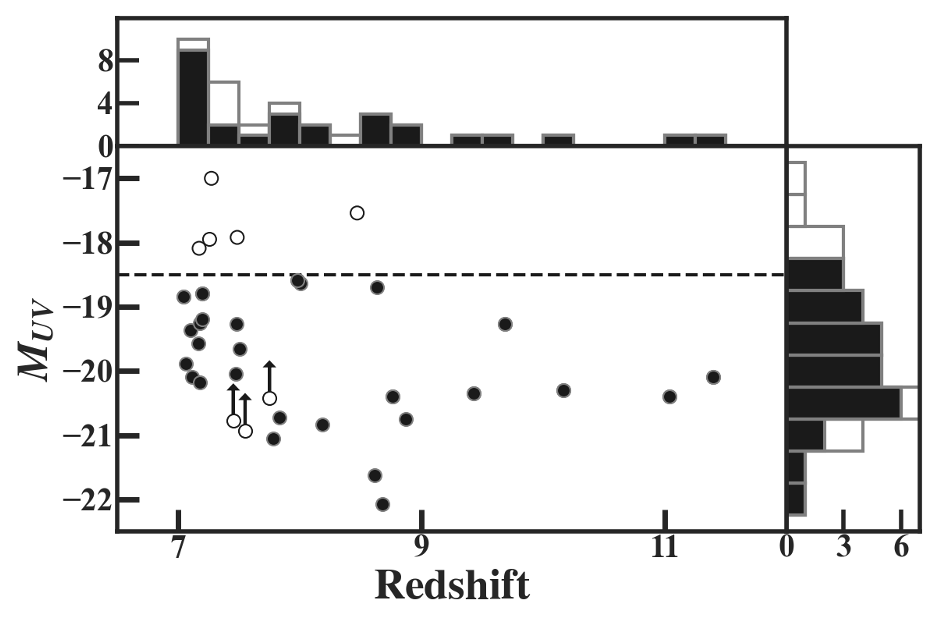}
\end{center}
\caption{Redshift and UV-magnitude distribution of our galaxy sample. \Add{In the bottom panel, circles represent the \Add{redshift and UV-magnitude} values for the galaxy observed by CEERS, GO-1433, DDT-2750, \Add{and JADES}. The circle indicate the best-fit values while the upper circle with an upper arrow corresponds to one sigma upper lower-limit for the UV-magnitudes. The filled circle indicates the galaxy used in our analysis, while the open circle indicates the galaxies omitted from our sample after a selection based on the UV-magnitudes. The top (right) panel shows the number counts at each redshift ($M_{\rm UV}$) bins for our galaxy sample. The fraction of bins represented by black represents the final sample after the absolution UV-magnitude selection.}}
\label{z_MUV}
\end{figure}

\begin{figure*}[t]
\centering
\begin{center}
\includegraphics[width=0.75\linewidth,clip]{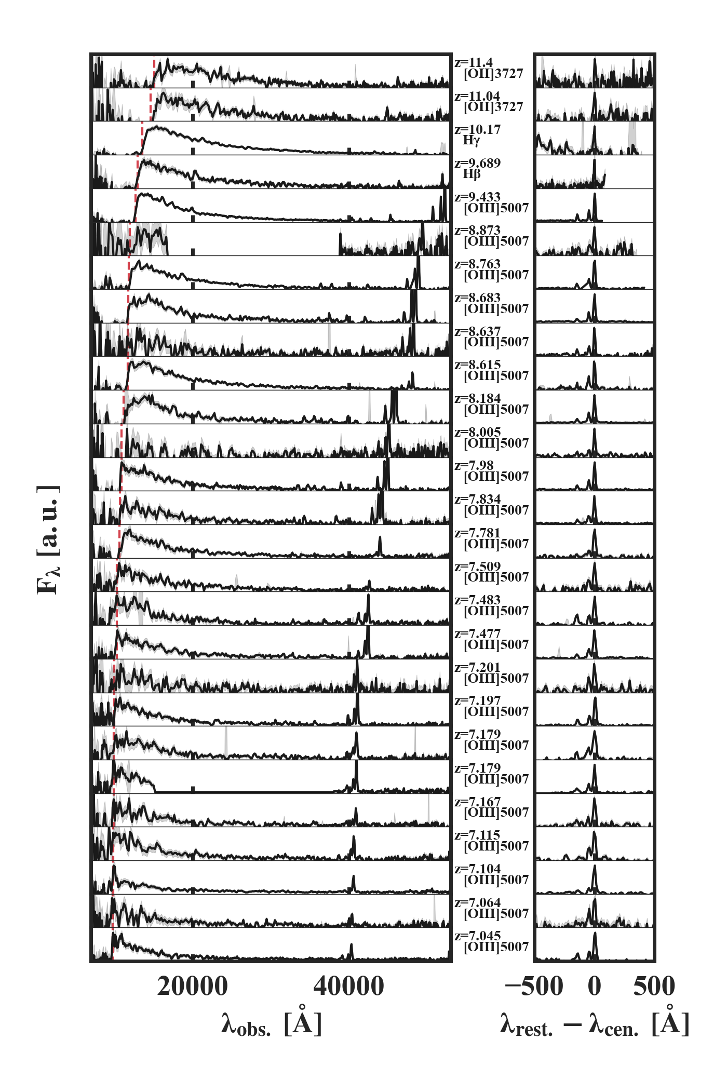}
\end{center}
\caption{Full spectra of galaxy samples. (Left): The x-(y-)axis corresponds to the observed wavelength (arbitrary normalized flux \Add{density}). The black lines and grey regions represent the fluxes of spectra and their uncertainties, respectively. The red dashed line represents the observed Ly$\alpha$ wavelength for each object. (Right) \Add{The highest $S/N$ emission lines detected from each spectra in our sample. The type of emission line shown and the name of the corresponding galaxy is shown between the right and left panels.} The x-axis represents the difference between the rest-frame wavelength $\lambda_{\rm rest.}$ and the central wavelength $\lambda_{\rm cen.}$ of the optical emission line.}
\label{full_spec}
\end{figure*}

We use spectroscopic datasets obtained in multiple observation programs; the Cosmic Evolution Early Release Science \citep[CEERS; ERS-1345. PI. S. Finkelstein;][]{Fi23}, GO-1433 \citep[PI. D. Coe;][]{H23}, DDT-2750 \citep[PI. P. Arrabal Halo;][]{AH23b}, \Add{and the JWST Advanced Deep Extragalactic Survey \citep[JADES; GTO-1210; PI. N. Lützgendorf;][]{Bun23}.} We only used the PRISM data ($R\sim100$) that covers $0.6-5.3~\mu$m. The total exposure time for the PRISM spectra from CEERS, GO-1433, DDT-2750, and JADES data are 0.86, 3.7, 5.1, \Add{and 9.3-27.7} hours, respectively. As presented in \cite{Nk23} and \cite{Ha23}, all the PRISM data from CEERS, GO-1433, and DDT-2750 we use are reduced with the JWST pipeline version 1.8.5 with the Calibration Reference Data System (CRDS) context file of {\tt jwst1028.pmap} or {\tt jwst1027.pmap} with some improvements on the flux calibration, noise estimate, and the composition. \Add{For the JADES data, we use the publicly available reduced data from JADES group \footnote{\url{https://archive.stsci.edu/hlsp/jades}}. Please see \cite{Bun23} for the detailed reduction processes of JADES data.} We then select the galaxies with systemic redshifts above $z>7$ confirmed with emission lines. \Add{For the $z<9$ galaxies in CEERS, we adopt the spectroscopic redshift reported by \cite{Nk23} in which the authors determine spectroscopic redshift via fitting \Add{{\OIII\W\W}4959,5007 and {\Hb}} triplet simultaneously. For the galaxies at $z>9$ obtained from observational program beside JADES, we adopt the spectroscopic redshift reported by \cite{Ha23}. In \cite{Ha23}, the authors derive the spectroscopic redshift for $z>11$ object via \Add{{\OII\W}3727 (and {\NeIII\W}3869)} for Maisie's galaxy (CEERS2\_588). For MACS0647\_JD, \cite{Ha23} use {\Hc} to determine spectroscopic redshift. For JADES objects, we adopt redshift reported in the public catalog released by the JADES team. The determination of the spectroscopic redshift listed in the public JADES is described in \cite{Bun23}. In \cite{Bun23}, the authors use emission lines detected with a signal to noise ratio of $S/N>5$ to determine the spectroscopic redshift. Although the uncertainty of the spectroscopic redshift is not given in the literature, \cite{Jon23} report that the spectroscopic redshift determined from PRISM data has a median offset of 0.00388 with a standard deviation of 0.00628. In Section \ref{syst}, we discuss on the effect of the uncertainty in the redshift confirmation with low resolution PRIMS data to our analysis.} Please see \cite{Nk23}, \cite{Ha23}, and \Add{\cite{Bun23}} for further details on the data reduction and spectroscopic redshift determination. We further narrow the sample by selecting galaxies with data available in wavelength corresponding to rest-frame $1200-1250$ {\AA} to investigate Ly$\alpha$ damping wing features. We obtain a total of \Add{35} galaxies as our galaxy sample candidates. Our sample candidates consist of \Add{35} galaxies from $z=$\Add{7.045} to 11.40\Add{,} with the redshift distribution shown in Figure \ref{z_MUV}. \Add{In Table \ref{table:gal}, we summarize properties of the sample candidate galaxies.} \Add{We apply additional sample selection by setting an absolute UV magnitude cut of $M_{\rm UV}<-18.5$ to omit faint ($M_{\rm UV}>-18.5$) galaxies distributed mainly at $z<8$ in our sample candidates. We also discard three of the object which \cite{Nk23} only report 5$\sigma$ upper limit value for UV magnitude for Hubble Space Telescope (HST) observations given that these objects are not detected by HST and lack JWST/NIRCam data.} After this magnitude cut selection, we have a final galaxy sample consisted by \Add{27} galaxies from $7.045<z<11.40$. Out of all \Add{27} galaxies, spectroscopic data of \Add{18}, 1, \Add{3}, and \Add{5} galaxies are obtained in CEERS, GO-1433, DDT-2750, and \Add{JADES} programs, respectively. We present \Add{the complete PRISM spectra} of galaxies in our \Add{final} sample in the left panels of Figure \ref{full_spec} in the order of the highest to lowest redshift galaxies. The black line and grey region in each panel represents the observed flux and the error of each galaxy spectra normalized by an arbitrary unit (a.u.), respectively. We also show the wavelengths of \Add{Ly$\alpha$} determined from the systemic redshifts by the red vertical dashed lines. Next to the spectra of each galaxy, we present \Add{the emission line detected in the PRISM spectra with the highest $S/N$ for} each galaxy in the right hand panels. The systemic redshift values and the name of the emission line shown for each galaxy are written in between the panels.\par

\begin{figure*}
\centering
\begin{minipage}{0.49\hsize}
\begin{center}
\includegraphics[width=0.95\hsize,clip]{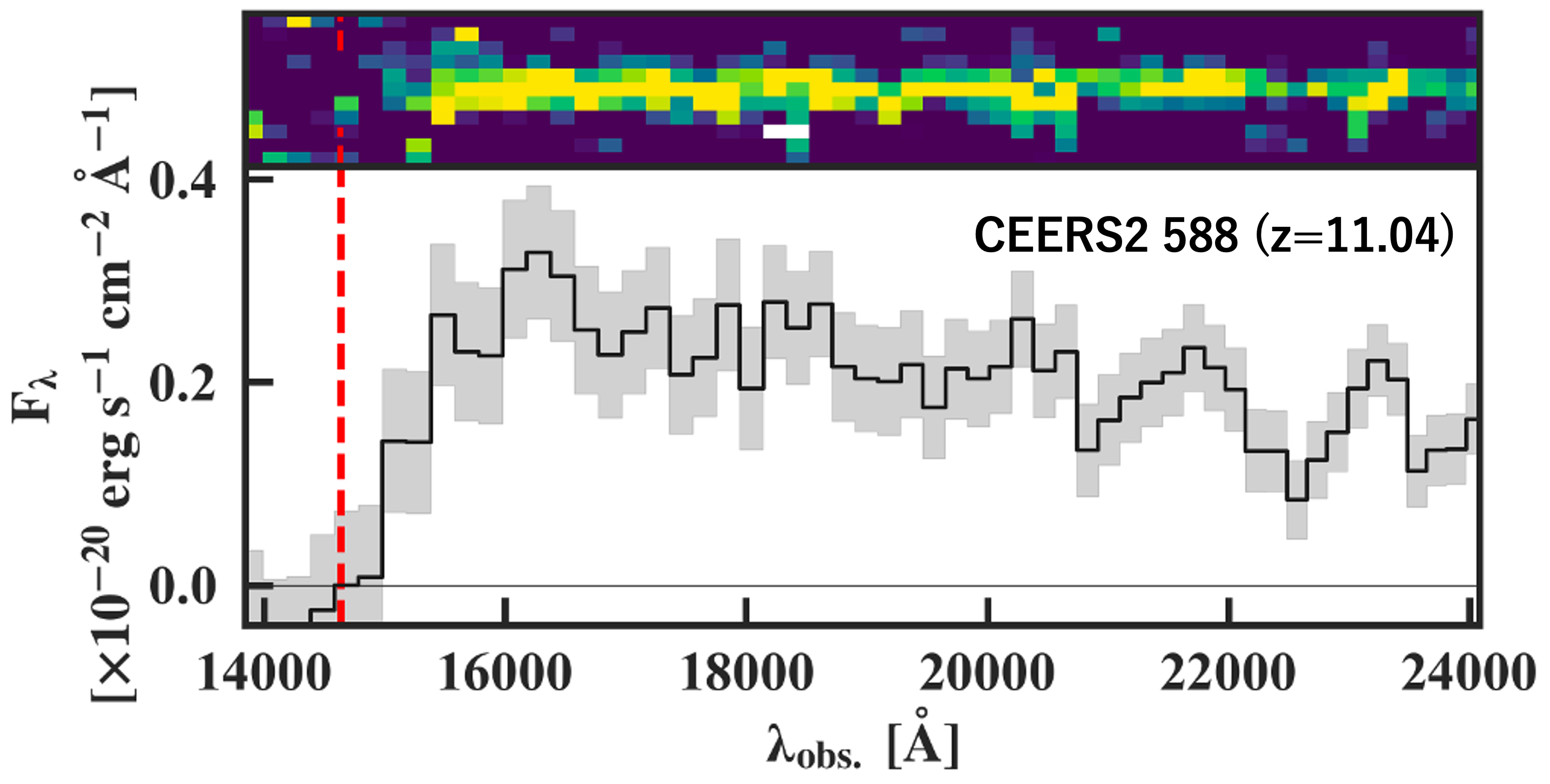}
\end{center}
\end{minipage}
\begin{minipage}{0.49\hsize}
\begin{center}
\includegraphics[width=0.95\hsize,clip]{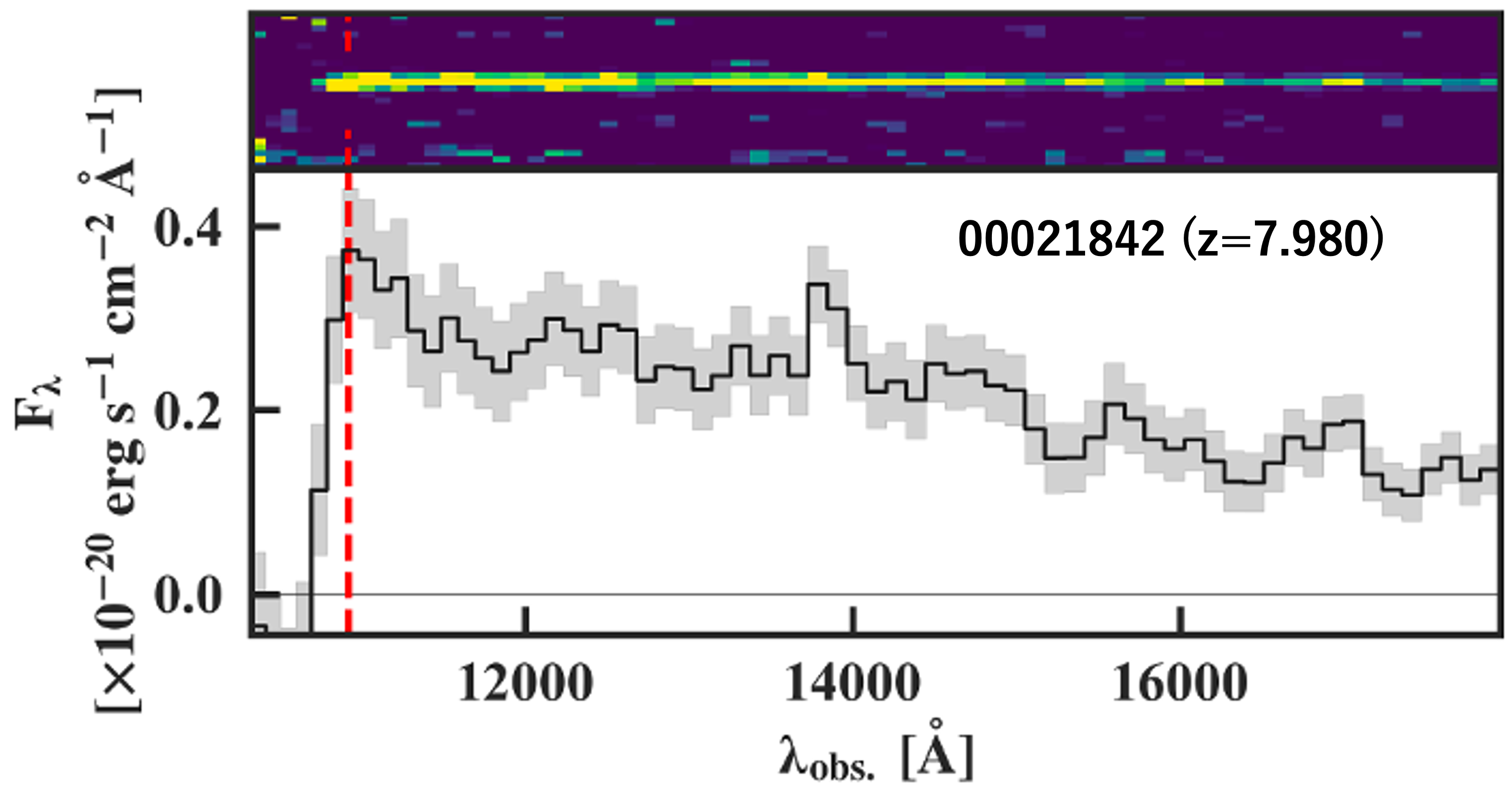}
\end{center}
\end{minipage}
\caption{\Add{2D (top) and 1D (bottom) spectra of CEERS2\_588 (left) and 00021842 (right). The black line and grey region represent the observed flux and its uncertainty, respectively. The red dashed line represents the wavelength of Ly$\alpha$ at the systemic redshift determined by non-Ly$\alpha$ emission lines.}}
\label{1D2D}
\end{figure*}

\Add{The spectra of the galaxy in our final sample show various features around rest-frame Ly$\alpha$ wavelength. We show two contrasting example in Figure \ref{1D2D}. The 2D and 1D spectra of CEERS2\_588 (00021842) with systemic redshift at 11.04 (7.980) is shown in the left (right) panel. The red dotted lines represent the systemic redshift determined by emission line. The spectra of CEERS2\_588 shows redshifted, softened Ly$\alpha$ break relative to systemic redshift, while we see blueshifted flux with 4.2$\sigma$ significance beyond the rest-frame 1216 {\AA} for 00021842. In Section \ref{res}, we physically interpret the diverse spectral features around the rest-frame 1216 {\AA} via spectral fitting.}\par

\begin{deluxetable}{cccccc}
\tablecolumns{6}
\tabletypesize{\scriptsize}
\tablecaption{Sample Properties%
\label{table:gal}}
\tablehead{%
\colhead{Name} & %Object?
\colhead{R.A.} &
\colhead{Decl.} &
\colhead{$z_{\rm spec}$} &
\colhead{$M_{\rm UV}$} &
\colhead{Ref.}\\
\colhead{} & %Object?
\colhead{(deg)} &
\colhead{(deg)} &
\colhead{} &
\colhead{(mag)} &
\colhead{}\\
\colhead{(1)} &
\colhead{(2)} &
\colhead{(3)} &
\colhead{(4)} &
\colhead{(5)} &
\colhead{(6)} 
}
\startdata
\Add{00020961} & 53.13423 & -27.76891 & 7.045 & -18.840 & l,m\\
CEERS\_00542 & 214.831624 & 52.831505 & 7.064 & -19.89 & h\\
CEERS\_00044 & 215.001115 & 53.011269 & 7.104 & -19.36 & h\\
CEERS\_00534 & 214.859117 & 52.85364 & 7.115 & -20.1 & h\\
CEERS\_00829 & 214.861594 & 52.876159 & 7.167 & -19.57 & h\\
CEERS\_80374 & 214.898074 & 52.824895 & 7.174 & -18.09 & h\\
CEERS\_00439 & 214.825364 & 52.863065 & 7.179 & -19.25 & h\\
CEERS\_00498 & 214.813045 & 52.834249 & 7.179 & -20.18 & h\\
\Add{10013905} & 53.11833 & -27.76901 & 7.197 & -18.800 & l,m\\
CEERS\_01038 & 215.039697 & 52.9015971 & 7.201 & -19.2 & h\\
\Add{00008079} & 53.15283 & -27.80194 & 7.260 & -17.950 & l,m\\
\Add{10013682} & 53.16746 & -27.77201 & 7.275 & -17.000 & l,m\\
CEERS\_01163 & 214.990468 & 52.9719902 & 7.455 & $>-20.78$ & h\\
CEERS\_80432 & 214.812056 & 52.746747 & 7.477 & -20.05 & h\\
CEERS\_80372 & 214.927798 & 52.850003 & 7.483 & -19.27 & h\\
CEERS\_80239 & 214.896054 & 52.869853 & 7.487 & -17.92 & h\\
CEERS\_80445 & 214.843115 & 52.747886 & 7.509 & -19.66 & h\\
CEERS\_00689 & 214.999053 & 52.9419767 & 7.552 & $>-20.93$ & h\\
CEERS\_00686 & 215.150862 & 52.9895618 & 7.752 & $>-20.42$ & h\\
CEERS\_01023 & 215.188413 & 53.0336473 & 7.781 & -21.06 & h\\
CEERS\_01027 & 214.882994 & 52.8404159 & 7.834 & -20.73 & h\\
\Add{00021842} & 53.15682 & -27.76716 & 7.980 & -18.590 & l,m\\
CEERS\_00003 & 215.005189 & 52.99658 & 8.005 & -18.64 & h\\
CEERS\_01149 & 215.089714 & 52.9661828 & 8.184 & -20.83 & h\\
\Add{00008013} & 53.16446 & -27.80218 & 8.473 & -17.540 & l,m\\
EGS\_z910\_44164 & 215.218762 & 53.0698619 & 8.615 & -21.63 & f,j \\
&&&&&h,d\\
CEERS\_90671 & 214.961276 & 52.842364 & 8.637 & -18.7 & b,d\\
CEERS\_1019 & 215.035391 & 52.8906618 & 8.683 & -22.07 & h,j,k \\
&&&&&i,g,d\\
\Add{CEERS\_43833} & 214.938625 & 52.911750 & 8.763 & -20.40 & c,h \\
CEERS1\_6059 & 215.011706 & 52.988303 & 8.873 & -20.75 & c,d,h \\
\Add{10058975} & 53.11243 & -27.77461 & 9.433 & -20.355 & l,m\\
\Add{00006438} & 53.16735 & -27.80750 & 9.689 & -19.270 & l,m\\	
MACS0647-JD${}^{\dag}$ & 101.982213 & 70.2432667 & 10.17 & -20.3 & d,e\\
CEERS2\_588${}^{\ddag}$ & 214.906625 & 52.9115278 & 11.04 & -20.4 & d\\
Maisie's Galaxy${}^{\ddag}$ & 214.943167 & 52.9424417 & 11.40 & -20.1 & a,d\\
(CR2-z12-1) & & & & & \\
\enddata
\tablecomments{
(1): Name.
(2): Right ascension of the object's coordinate.
(3): Declination of the object's coordinate.
(4): Spectroscopic redshift.
(5): Absolute UV magnitude.
(6): References for spectroscopic redshifts and UV magnitude:a.) \cite{AH23a}, b.) \cite{AH23b}, c.) \cite{Fu23}, d.) \cite{Ha23}, e.) \cite{H23}, f.)\cite{La22}, g.) \cite{La23}, h.) \cite{Nk23}, i.) \cite{Sa23}, j.) \cite{Tan23}, k.) \cite{Zi15}, \Add{l.) \cite{Bun23}, m.) \cite{Jon23}}\\
All spectroscopic data are from CEERS program except for the spectroscopic data of the object name with ${}^{\dag}$ and ${}^{\ddag}$ are from GO-1433 and DDT-2750 programs, respectively.
}
\end{deluxetable}

\section{Methods} \label{met}
\subsection{Ly${\alpha}$ Damping Wing} \label{dwfit}
To calculate the optical depth \Add{associated with Ly$\alpha$ damping wing absorption}, we assume that galaxy resides at the center of the fully ionized bubble with radius $R_b$ and the surrounding IGM with a constant $x_{\rm HI}$ until the end of EoR, $z_{n}$. We fix \Add{the redshift corresponding to the end of reionization at $z_n=5$} in our calculations. \Add{Note that changing $z_n$ by $\pm1$ have a negligible effect on IGM Ly$\alpha$ damping wing absorption profile.}\par
\begin{figure}[htbp]
\centering
\begin{center}
\includegraphics[width=\linewidth]{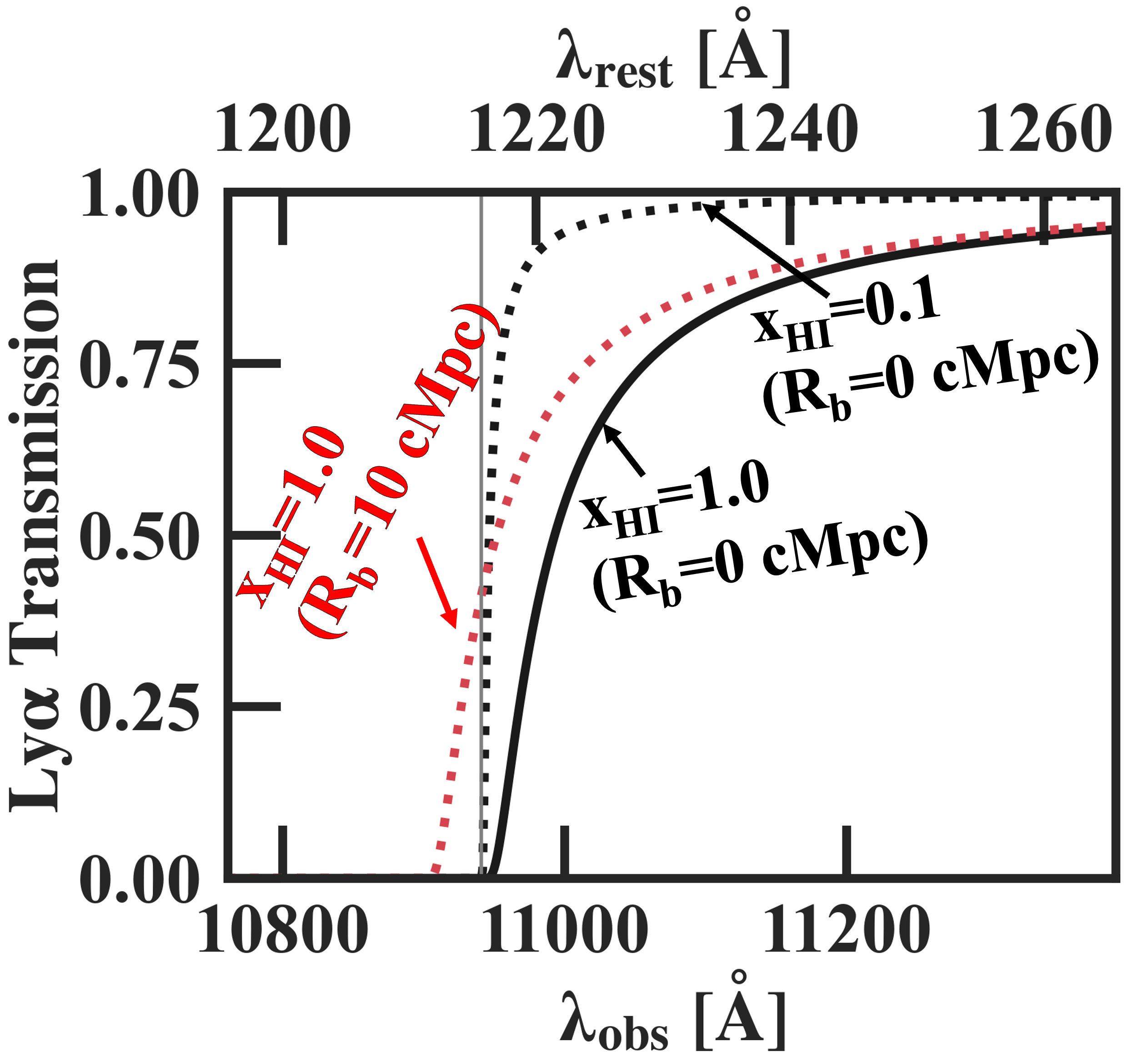}
\caption{Example of Ly$\alpha$ damping wing profile for $z=8$. The top and bottom axes are the rest frame and observed wavelengths, respectively. The y-axis corresponds to the transmission fraction of the flux at each wavelength. The solid (dotted) black line corresponds to the absorption profile with $x_{\rm \HI}=1.0$ (0.1) and $R_{b}=0$ cMpc. The red dotted line corresponds to the absorption profile with $x_{\rm \HI}=1.0$ and $R_{b}=10$ cMpc. The gray vertical line corresponds to the rest-frame Ly$\alpha$ wavelength.}
\label{example}
\end{center}
\end{figure}

To calculate the optical depth of Ly$\alpha$ damping wing at each observed wavelength \Add{$\lambda_{\rm obs}$}, $\tau(\lambda_{\rm obs})$, we adopt the equation 2 of \cite{2006PASJ...58..485T} that is based on the formulation of \cite{ME98}:
\begin{align}
\tau(\lambda_{\rm obs})&=\frac{x_{\rm \HI}\Lambda_{\alpha}\lambda_{\alpha}\tau_{0}(z_s)}{4\Add{\pi^2}c}\left(\frac{1+z_{\rm obs}}{1+z_{s}}\right)^{3/2}\\
&\times\left[I\left(\frac{1+z_{b}}{1+z_{\rm obs}}\right)-I\left(\frac{1+z_{n}}{1+z_{\rm obs}}\right)\right], \nonumber
\label{DW_eq}
\end{align}
where $\Lambda_{\alpha}$, $\lambda_{\alpha}$, $c$, and $\tau_0$, are the decay constant for the Ly\Add{$\alpha$} resonance, the rest-frame Ly\Add{$\alpha$} wavelength, the speed of light, and the Gunn-Peterson optical depth, respectively. \Add{$z_{b}$ is defined as} the redshift corresponding to \Add{the edge of the} ionized bubble with radius $R_b$, whereas $z_{\rm obs}$ is a value defined as $\lambda_{\rm obs}/\lambda_\alpha-1$. $I(x)$ is \Add{an integration defined as follows}:
\begin{align}
    I(x)&=\frac{x^{9/2}}{1-x}+\frac{9}{7}x^{7/2}+\frac{9}{5}x^{5/2}+3x^{3/2} \\
    &+9x^{1/2}-\frac{9}{2}\ln{\frac{1+x^{1/2}}{1-x^{1/2}}}. \nonumber
\end{align}
We calculate $\tau_0$ using the following formula also given by \cite{ME98}:
\begin{equation}
\tau_0(z_s)=\frac{3\lambda_\alpha^3\Lambda_\alpha n_0}{8\pi H(z_s)},
\end{equation}
where $n_0$ and $H(z)$ are the hydrogen number density and the Hubble \Add{parameter} as the function of redshift. We calculate $n_0$ using the critical density of the universe, \Add{baryon density parameter, and primordial helium abundance}. We show how the absorption profile of the $z=8$ changes with different $x_{\rm HI}$ and $R_b$ values in Figure \ref{example}. As shown in Figure \ref{example}, increasing $x_{\rm HI}$ value extends the absorption feature to a redder side of 1216 \AA\ while increasing $R_{b}$ value shifts the absorption profile to the bluer wavelength. Notably, fluxes on the bluer side of the 1216 \AA\ can be transmitted if the bubble size is sufficiently large (i.e., $R_b>10$ comoving Mpc; cMpc).\par

\subsection{Stacking Spectra}
 We construct composite spectra by stacking individual spectra to maximize the signal to noise ratio. We divide the galaxy sample into four subsamples by binning according to the redshifts so that each subsample has at least six galaxies. We label the subsamples as \#1, \#2, \#3, and \#4 bins in the order of the lower to higher redshift. The redshift range for \#1, \#2, \#3, and \#4 bins are \Add{7.045-7.179, 7.197-7.781, 7.834-8.683, and }\Add{8.763}-11.40, respectively. We define representative redshift values $\langle z \rangle$ of each redshift bin by the average redshift value among the galaxies in each subsample.\par
\begin{figure*}[htbp]
\centering
\begin{center}
\includegraphics[width=\linewidth]{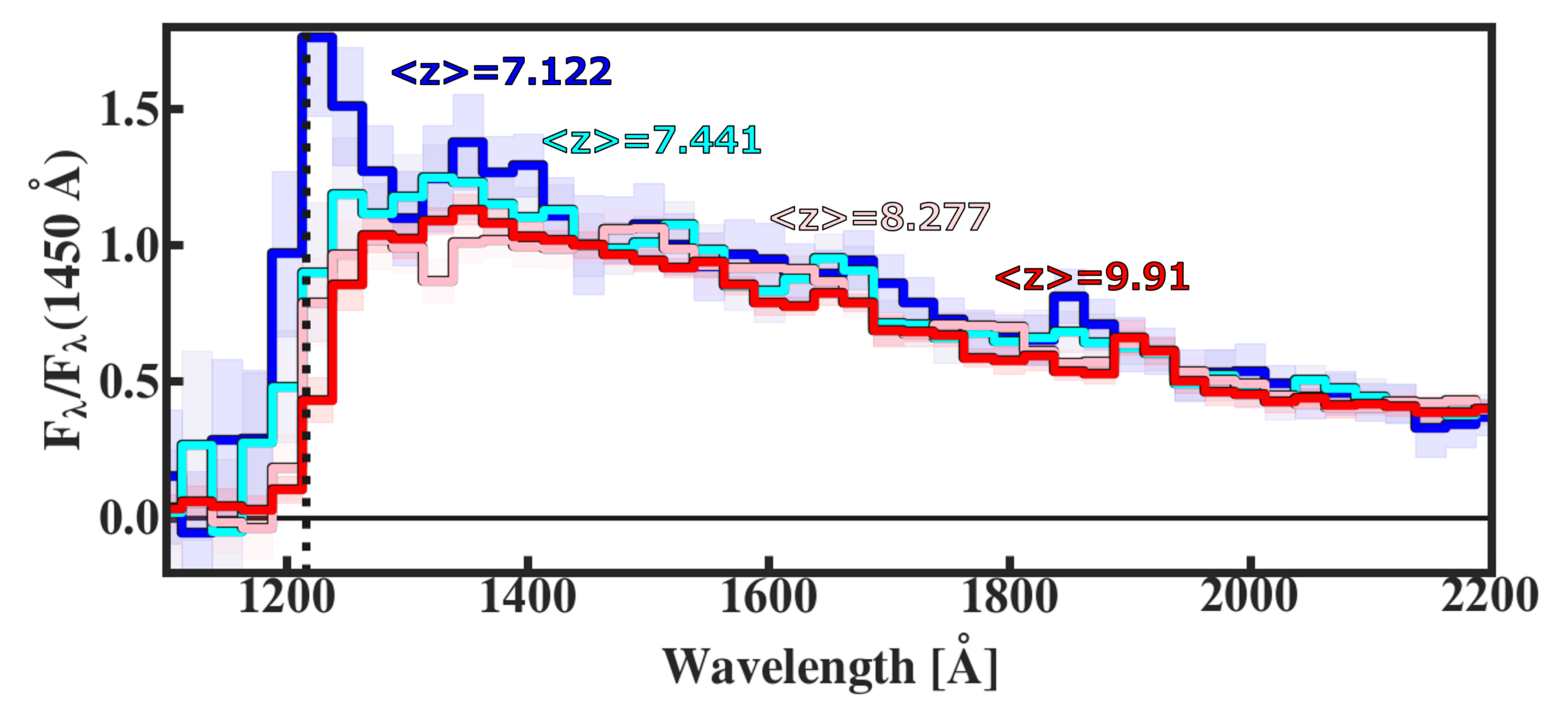}
\end{center}
\caption{Stacked spectra of our galaxy sample binned into four redshift bins. The \Add{dark blue, pale blue, pink, and red solid lines (shades)} represent the stacked spectra (uncertainties) for the redshift bins at \Add{$\langle z \rangle=$7.12, 7.44, 8.28, and \Add{9.91}}, respectively. The wavelength is shifted to the rest frame and the flux is normalized at \Add{1450} {\AA} \Add{at which the IGM absorption have negligible impact on the spectrum}. \Add{The black vertical dotted line represent the position of the rest-frame 1216 {\AA}. The horizontal solid line indicates the position of the flux value at zero.}}
\label{stack}
\end{figure*}

 For stacking individual spectra of each subsample, we have shifted the galaxy spectra into rest-frame wavelength based on the systemic redshift and resampled to a common wavelength pixels with a pixel size of 25 \AA \Add{, a value considerably similar to that of the resolution of PRISM at $<$2 $\mu$m ($R\sim30$),} using linear interpolation. We have taken error weighted \Add{mean} flux values at wavelength pixel to calculate the stacked spectra. We estimate the uncertainty of the stacked spectra via 10000 iterations of bootstrap resampling. During each resampling, we fluctuate the observed flux at each pixel with Gaussian noise accordingly to the original error spectrum to consider the measurement error as well. We present the stacked spectra from all subsamples in Figure \ref{stack}. Fluxes of the composite spectra are normalized at the rest frame \Add{1450} \AA. As shown in Figure \ref{stack}, we can clearly see the \Add{softening break feature} at the rest-frame 1216 {\AA} towards the high redshift, suggesting increasing Ly$\alpha$ damping wing absorptions due to increasing $x_{\rm \HI}$ with redshifts.  
\begin{deluxetable}{ccccc}
\tablecolumns{5}
\tabletypesize{\scriptsize}
\tablecaption{Subsamples Properties%
\label{table:sub}}
\tablehead{%
\colhead{ID} & %Object?
\colhead{$N$} &
\colhead{$\langle z \rangle$} &
\colhead{$z_{\rm lower}$} &
\colhead{$z_{\rm upper}$} \\
\colhead{(1)} &
\colhead{(2)} &
\colhead{(3)} &
\colhead{(4)} &
\colhead{(5)}
}
\startdata
\#1 & 7 & 7.12 & 7.045 & 7.179\\
\#2 & 6 & 7.44 & 7.197 & 7.781\\
\#3 & 7 & 8.28 & 7.834 & 8.683 \\
\#4 & \Add{7} & \Add{9.91} & \Add{8.763} & 11.40 \\
\enddata
\tablecomments{
(1): ID for a subsample.
(2): Size of a subsample.
(3): Representative (i.e., the average) redshift of galaxies in a subsample.
(4): Lowest redshift of galaxies in a subsample.
(5): Highest redshift of galaxies in a subsample.
}
\end{deluxetable}

\subsection{Spectral Fitting} \label{cfit}

\begin{deluxetable*}{cccccccccc}
\tablecolumns{10}
\tabletypesize{\scriptsize}
\tablecaption{Prior Distributions for Fitting Parameters%
\label{table:prior}}
\tablehead{%
\colhead{$\log M_{\star}$} &
\colhead{$\log(SFR_{i+1}/SFR_{i})$} &
\colhead{$\tau_{5500 {\rm \AA}}$} &
\colhead{$\log U$} &
\colhead{${x_{\rm \HI}}$} &
\colhead{$\log R_b$} &
\colhead{$\log N_{\rm \HI}$} &
\colhead{$\Delta_{v, \rm DLA}$} &
\colhead{$\sigma_{v}$} &
\colhead{$\log EW_{\alpha,0}$} \\
\colhead{$(M_{\odot})$} &
\colhead{$(i=1,2,3,4,5)$} &
\colhead{} &
\colhead{} &
\colhead{} &
\colhead{(cMpc)} &
\colhead{(${\rm cm}^{-2}$)} &
\colhead{$({\rm km~s^{-1}})$} &
\colhead{$({\rm km~s^{-1}})$} &
\colhead{$({\rm \AA})$} \\
\colhead{(1)} &
\colhead{(2)} &
\colhead{(3)} &
\colhead{(4)} &
\colhead{(5)} &
\colhead{(6)} &
\colhead{(7)} &
\colhead{(8)} &
\colhead{(9)} &
\colhead{(10)} 
}
\startdata
U(6,11) & 
t(0,0.3,2) & 
U(0,2) & 
U(-6,2) & 
U(0,1) & 
U(-1,3) & 
U(10,23) & 
N(0,300) & 
U(0,100) & 
U(-3,3) 
\enddata
\tablecomments{
(1): Total stellar mass. 
(2): Star formation rate ratio for six star formation history bin. 
(3): Optical depth at 5500 {\AA} for dust attenuation law by \cite{2000ApJ...533..682C}. 
(4): Ionization parameter. 
(5): Neutral hydrogen fraction.
(6): Ionized bubble radius.
(7): Neutral hydrogen column density for CGM.
(8): Velocity offset for CGM relative to galaxy systemic redshift.
(9): Velocity dispersion for CGM.
(10): Intrinsic Ly$\alpha$ equivalent width.\\
N($\mu$, $\sigma)$ is a normal distribution with mean $\mu$ and variance $\sigma^2$. U($a$, $b$) is a uniform distribution between $a$ and $b$. t(${\bm \mu}$, ${\bm \sigma}$, ${\bm \nu}$) is a student's t distribution with mean $\mu$, scale factor $\sigma$, and degree of freedom $\nu$.
}
\end{deluxetable*}

\Add{To further confirm the redshift evolution of Ly$\alpha$ damping wing absorption, we measure the IGM Ly$\alpha$ damping wing absorption. At the same time, we also consider the effect on the shape of Lyman break from the intrinsic galaxy's property such as stellar mass. \Add{Because the interpretation of spectral fitting using stacked spectrum would be complicated as spectra of galaxies with different physical properties are mixed together, we fit individual galaxy’s spectrum individually and consider marginalized posterior distributions for damping wing parameters.} For the fitting, we use the observed spectra in the wavelength range above the rest-frame 1000 {\AA} \Add{without masking Ly$\alpha$ emission}. We included data at shorter wavelength value than the rest-frame Ly$\alpha$ (i.e., 1216 \AA) to consider the effect of higher flux transmission due to large surrounding ionized bubble. \Add{We generate model spectrum in forward modeling manner} to compare the observed spectrum to model spectrum. We first generate the model spectra incorporating stellar and nebular emission. \Add{We also add Ly$\alpha$ emission prior to applying the Ly$\alpha$ absorption. Then, we apply Ly$\alpha$} absorption \Add{to the intrinsic model spectrum}. After \Add{applying Ly$\alpha$ absorptions}, we smooth the model spectra with line spread function of PRISM provided by STScI to account for the instrumental broadening. \footnote{\url{https://jwst-docs.stsci.edu/jwst-near-infrared-spectrograph/nirspec-instrumentation/nirspec-dispersers-and-filters}}}\par
\Add{We use a SED fitting code based on {\tt prospector} \citep{2021ApJS..254...22J} to generate model galaxy spectra with various stellar population and nebular properties. We use the Binary Population and Spectral Synthesis \citep[BPASS;][]{2017PASA...34...58E} model as an isochrone library. We allow different non-parametric star formation history (SFH), stellar mass ($M_{\star}$), ionization parameter ($\log U$), and dust attenuation. We adopt the star formation history in the same manner as \cite{Ha23}. We also allow different stellar mass in the range of $6<M_{\star}<11$, which fully covers the mass range of $z>7$ galaxies estimated in \cite{Nk23} and \cite{Ha23}. For the dust attenuation, we apply the attenuation law by \cite{2000ApJ...533..682C} and use an optical depth at 5500 {\AA} as a free parameter. We also include nebular emission continuum and lines, but we treat Ly$\alpha$ emission line separately as we describe later in this section. We fix stellar/nebular metallicity to 10\% solar metallicity as suggested for JWST high-z galaxy by recent work of \cite{Nk23}. \Add{We set the same prior distributions for free stellar/nebular parameters as adopted in \cite{Nk23}}.} \par 
\Add{For Ly$\alpha$ absorption, we consider Ly$\alpha$ absorption by both IGM and circum-galactic medium (CGM). \Add{We define the intrinsic Ly$\alpha$ emission as the Ly$\alpha$ emission before encountering CGM/IGM absorption. For intrinsic Ly$\alpha$ emission, we let the equivalent width of the intrinsic Ly$\alpha$ emission line ($\log EW_{\alpha,0}$) free in the range of $-3<\log EW_{\alpha,0}/{\rm \AA}<3$. In this way,  we can consider both extreme and negligible (i.e., low escape fraction) contribution of Ly$\alpha$ emission to the shape of Lyman break. We discuss about the validity of Ly$\alpha$ emission line strengths in Section \ref{fesc_disc}. We fix the velocity offset and dispersion of the Ly$\alpha$ emission line to +200 km/s and 250 km/s, respectively. We also discuss about how this assumption on the velocity offset affect the result in Section \ref{syst}.} \Add{For Ly$\alpha$ absorptions, w}e adopt the formulation given by equation \ref{DW_eq} for the IGM absorption with $x_{\rm \HI}$ and $\log R_{b}$ as free parameter. For the {\HI} absorption by CGM, we consider the Voigt profile with different neutral hydrogen column density ($\log N_{\rm \HI}$), velocity dispersion ($\sigma_v$), and velocity offset relative to the systemic redshift ($\Delta_{v, \rm DLA}$). \Add{We summarize our fitting parameters and corresponding prior distributions in Table \ref{table:prior}.}}\par
\Add{We run Markov Chain Monte Carlo (MCMC) method to obtain the posterior probability distribution functions (PDFs) of free parameters. \Add{To take an account of variance in galaxy's intrinsic spectral properties in estimating Ly$\alpha$ absorption parameters, we simultaneously fit for the stellar, nebular, and absorptions.} We use a package created based on {\tt emcee} \citep{2013PASP..125..306F} called {\tt ptemcee} \citep{2021ascl.soft01006V} for MCMC. {\tt ptemcee} incorporates parallel tempering technique that realizes faster sampling convergence in case of multi-modal posterior probability distribution functions. We run four temperature parallel tempering MCMC sampler of 80 walkers for each temperature with 5000 steps. We discard first 1000 steps of each walker to diminish the dependence of posterior PDFs to initial parameter distributions.}\par
\Add{We infer the best-fit parameters from the posterior PDFs. Because a median value does not well represent the skewed posterior distribution, we define the best-fit parameter by the mode (i.e., a peak of the posterior distribution) and uncertainty by 68th percentile highest posterior density interval (HPDI; i.e., a narrowest interval containing 68\%) of the posterior PDFs. To obtain mode of 1D marginalized posterior PDF for each parameter, we smooth the distribution by Kernel Density Estimation smoothing using {\tt scipy} package \citep{2020SciPy-NMeth} and choose the mode within an HPDI of 1D posterior distribution. We also \Add{calculate} the posterior probability distribution within each redshift bin (hereafter, \Add{binned posterior PDF) in the following manner:}}

\begin{equation}
\begin{aligned}
p(\mathbf{\theta}|\mathbf{D})&=
\frac{p(\mathbf{D}|\mathbf{\theta})p(\mathbf{\theta})}{p(\mathbf{D})} \\
&=\left[ \prod_{i}^{N}\frac{p(D_i|\mathbf{\theta})p(\mathbf{\theta})}{p(D_i)} \right]p^{1-N} (\mathbf{\theta}) \\
&=\left[ \prod_{i}^{N}p(\mathbf{\theta}|D_i) \right]p^{1-N} (\mathbf{\theta}).
\end{aligned}
\label{bayes}
\end{equation}
\Add{Here, $p(\cdot)$, $\mathbf{\theta}$, \Add{$N$,} and \Add{$D_i$ ($\mathbf{D}\equiv\left\{D_i\right\}_{i=1,...,N}$)} represent the probability density, free parameters, \Add{number of galaxy in each redshift bin}, and \Add{$i$-th (all set of)} galaxy spectral data in each bin, respectively. In equation \ref{bayes}, we assume}
\begin{equation}
\begin{aligned}
p(\mathbf{D})&=\prod_{i}^{N}p(D_i),\\
p(\mathbf{D}|\mathbf{\theta})&=\prod_{i}^{N}p(D_i|\mathbf{\theta}).
\end{aligned}
\label{bayes_assume}
\end{equation}
\Add{We utilize the posterior PDF for individual galaxies (i.e., $p(\mathbf{\theta}|D_i)$) obtained with MCMC to calculate the binned posterior PDFs.}
We obtain the best-fit value \Add{for each redshift bin in the similar manner as for the individual galaxies}.

\subsection{Fitting Results} \label{res}
\begin{figure*}
\centering
\begin{minipage}{0.49\hsize}
\begin{center}
\includegraphics[width=0.99\hsize,clip]{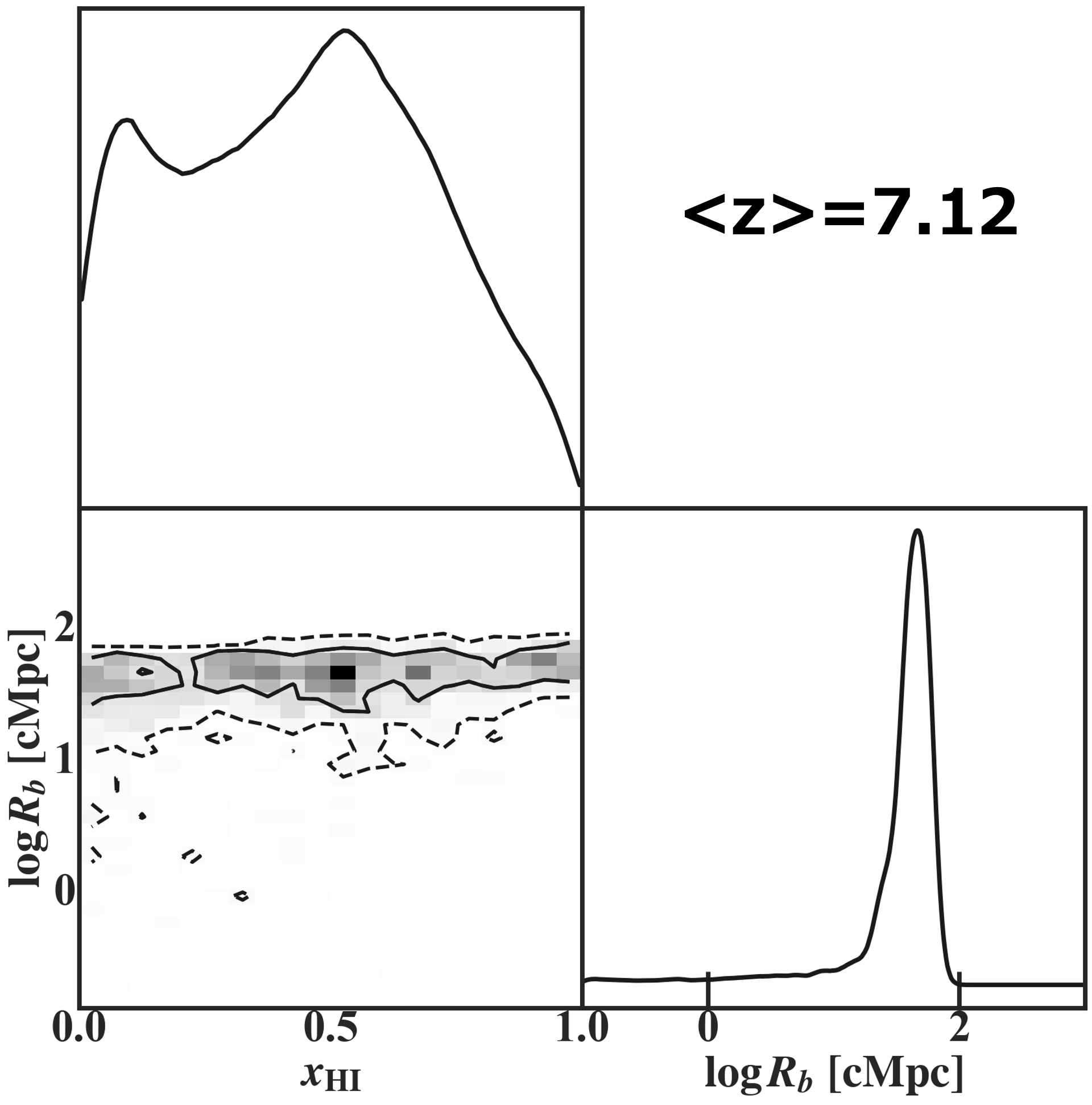}
\end{center}
\end{minipage}
\begin{minipage}{0.49\hsize}
\begin{center}
\includegraphics[width=0.99\hsize,clip]{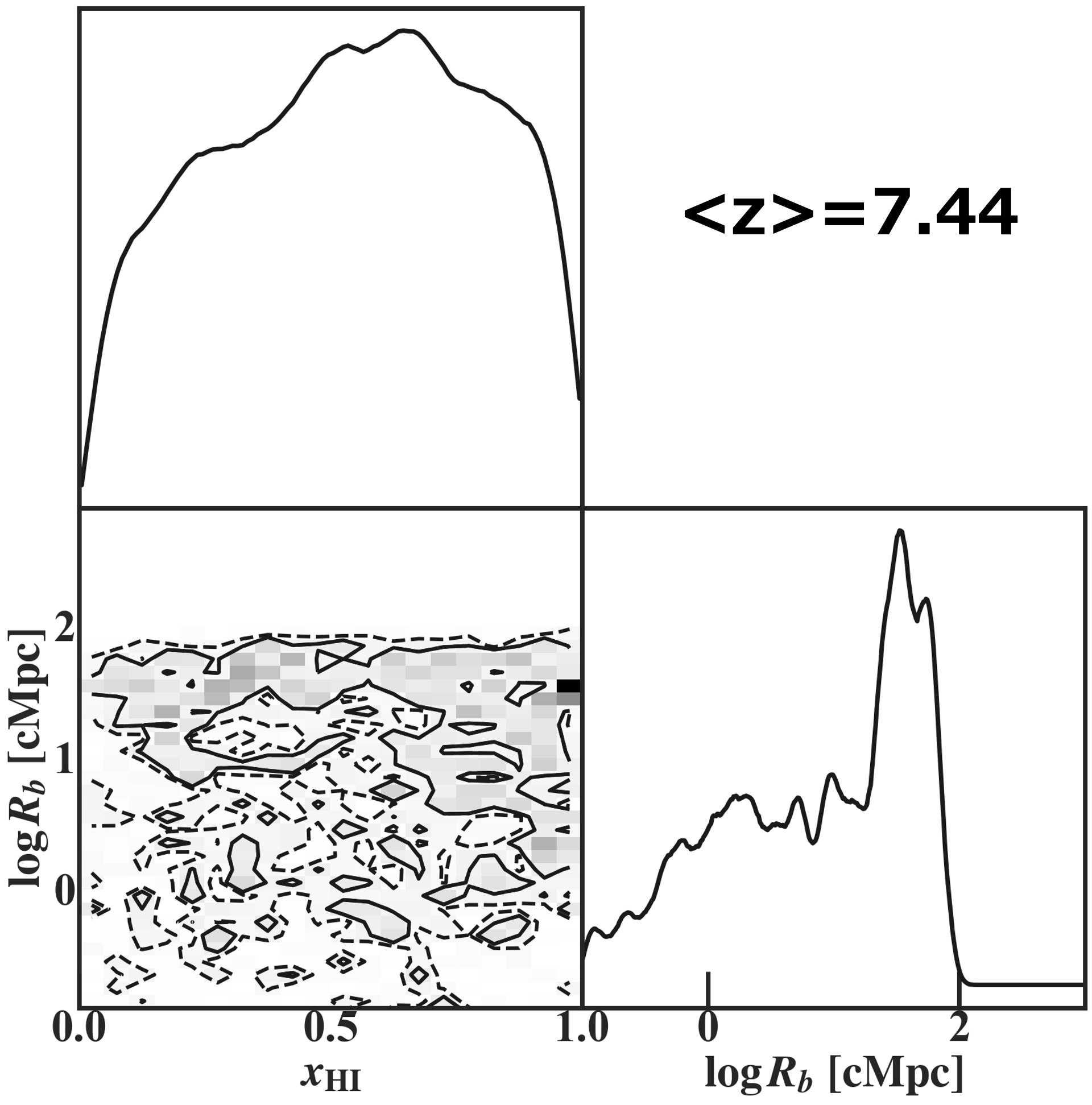}
\end{center}
\end{minipage}

\par
\vspace{0.5cm}

\begin{minipage}{0.49\hsize}
\begin{center}
\includegraphics[width=0.99\hsize,clip]{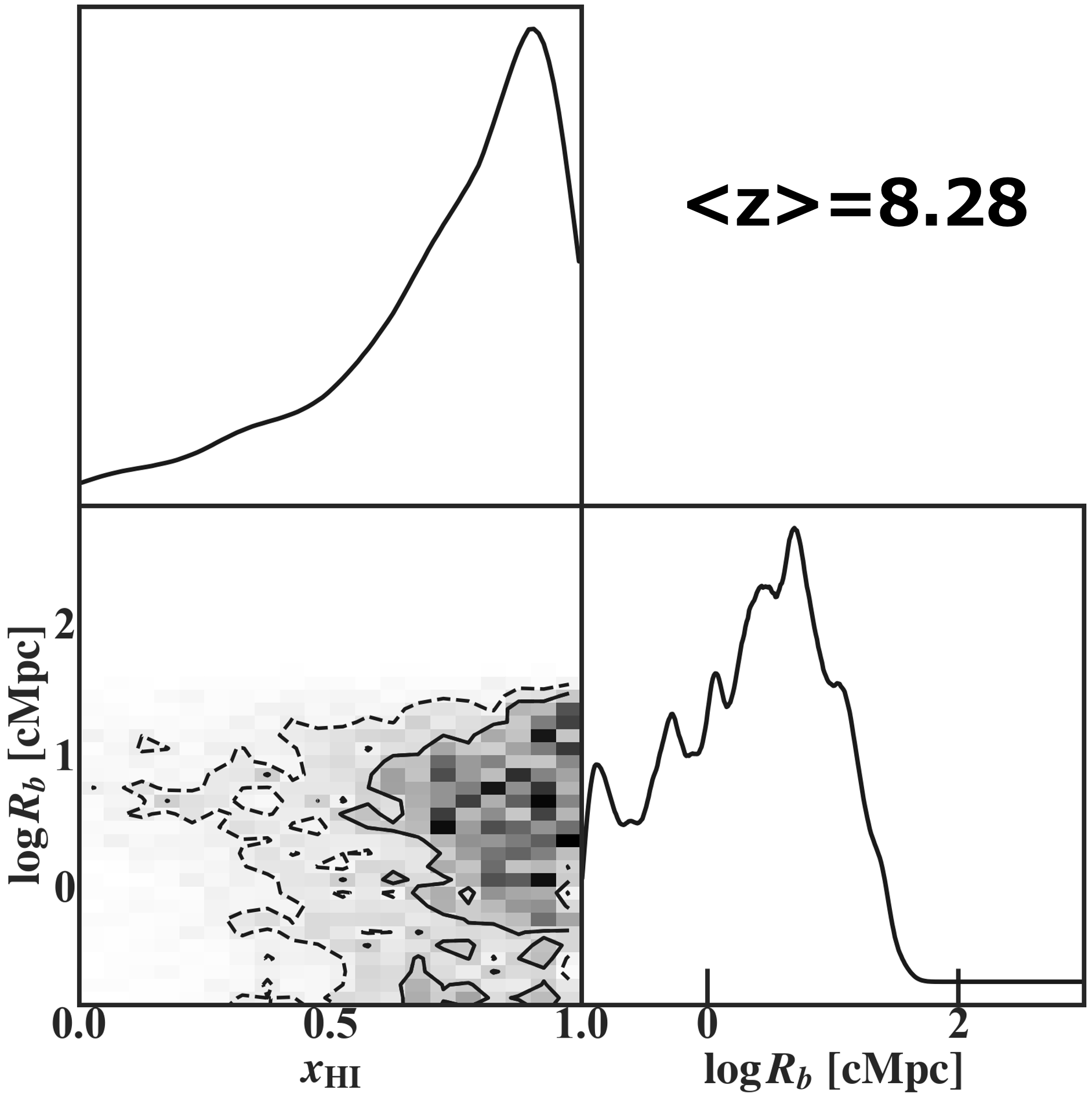}
\end{center}
\end{minipage}
\begin{minipage}{0.49\hsize}
\begin{center}
\includegraphics[width=0.99\hsize,clip]{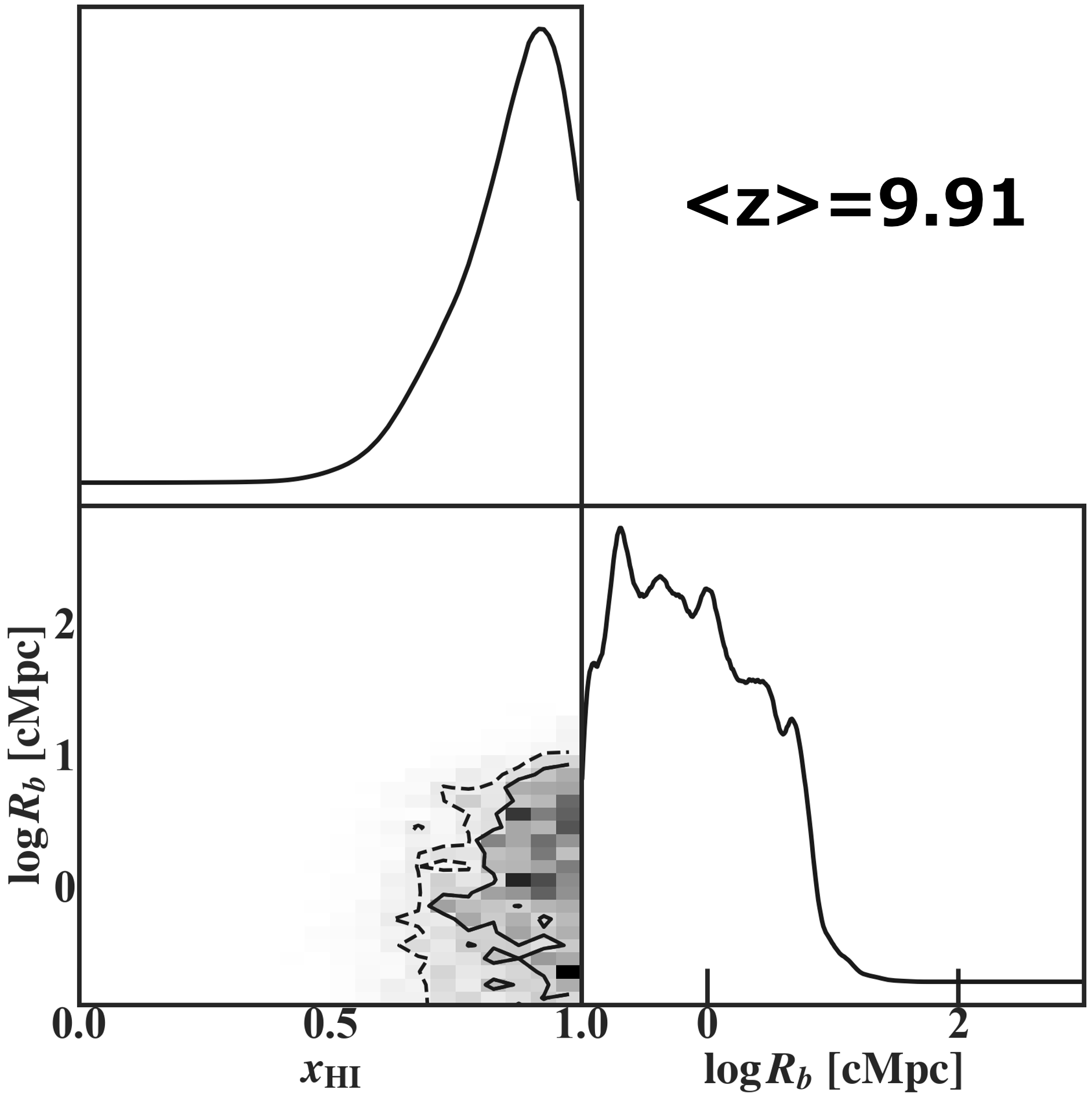}
\end{center}
\end{minipage}

\caption{\Add{The posterior probability distribution function (PDF) of \Add{$x_{\rm \HI}$ and $\log R_b$} for each subsample. The results for \#1  to \#4 bin are shown in order of top left to right bottom. \Add{The darkness in the 2D-marginalized PDFs represent the probability density. The solid and dotted contours represent the 68 and 90-th percentiles, respectively.} \Add{Top panel corresponds to the 1D marginalized PDF for each parameter. \Add{The presented PDFs are smoothed by gaussian kernel.}}}}
\label{res:fit_res}
\end{figure*}

\Add{In Figure \ref{res:fit_res}, we show the \Add{binned} posterior PDF of the parameter of ${x_{\rm \HI}}$ and $\log R_b$ for each redshift bin.} \Add{The solid (dashed) contour represents the 68-(90-)th percentile range for 2D marginalized PDFs.} The \Add{top two panels} in corner diagrams represent the marginalized probability density function of the $x_{\rm \HI}$ and \Add{$\log R_{b}$}. We summarize the best-fit parameters in Table \ref{table:bf}. For the results of the \#4 bin, we obtain very high $x_{\rm \HI}$ and small $R_b$ values (i.e., $x_{\rm \HI}\sim1$, $R_b~\Add{<1}$ cMpc) as similar as the ones reported for a $z=10.17$ galaxy in \cite{H23}. \Add{For the $z\sim7$ galaxies, our $x_{\rm \HI}$ is not well constrained as one can see from the marginalized PDF. The reason for tighter constraint on $\log R_b$ compared to that for $x_{\rm \HI}$ could be attributed to the strong dependence of $\log R_b$ value to the position of observed Lyman break relative to the systemic redshift.} Moreover, the reasons for having sharp peaks for $\log R_b$ compared with that for $x_{\rm \HI}$ could be attributed to that Ly$\alpha$ damping wing shapes are more sensitive $\log R_b$ than for $x_{\rm \HI}$ as $\log R_b$ have uniform prior distributions based on logarithmic form, unlike the linearly uniform prior distributions for $x_{\rm \HI}$. \Add{Logarithmically changing $\log R_b$ values have a little impact on the shape of model spectra, in another word the likelihood values do not change dramatically, when the $\log R_b$ is considerably small. This is reflected on the low and flat posterior probability values at small $\log R_b$ range and sharp peaks around the best-fit values in Figure \ref{res:fit_res}. We also confirm similar behaviors for posterior probability distribution for $\log R_b$ for individual galaxies. We note that narrowing down the range of flat priors for $\log R_b$ does not affect the best-fit value as it is defined by the mode, but the lower bound for uncertainty would increase if we adopt higher value for the lower bound of the flat prior.} \par

\begin{figure*}
\centering
\includegraphics[width=0.8\hsize,clip]{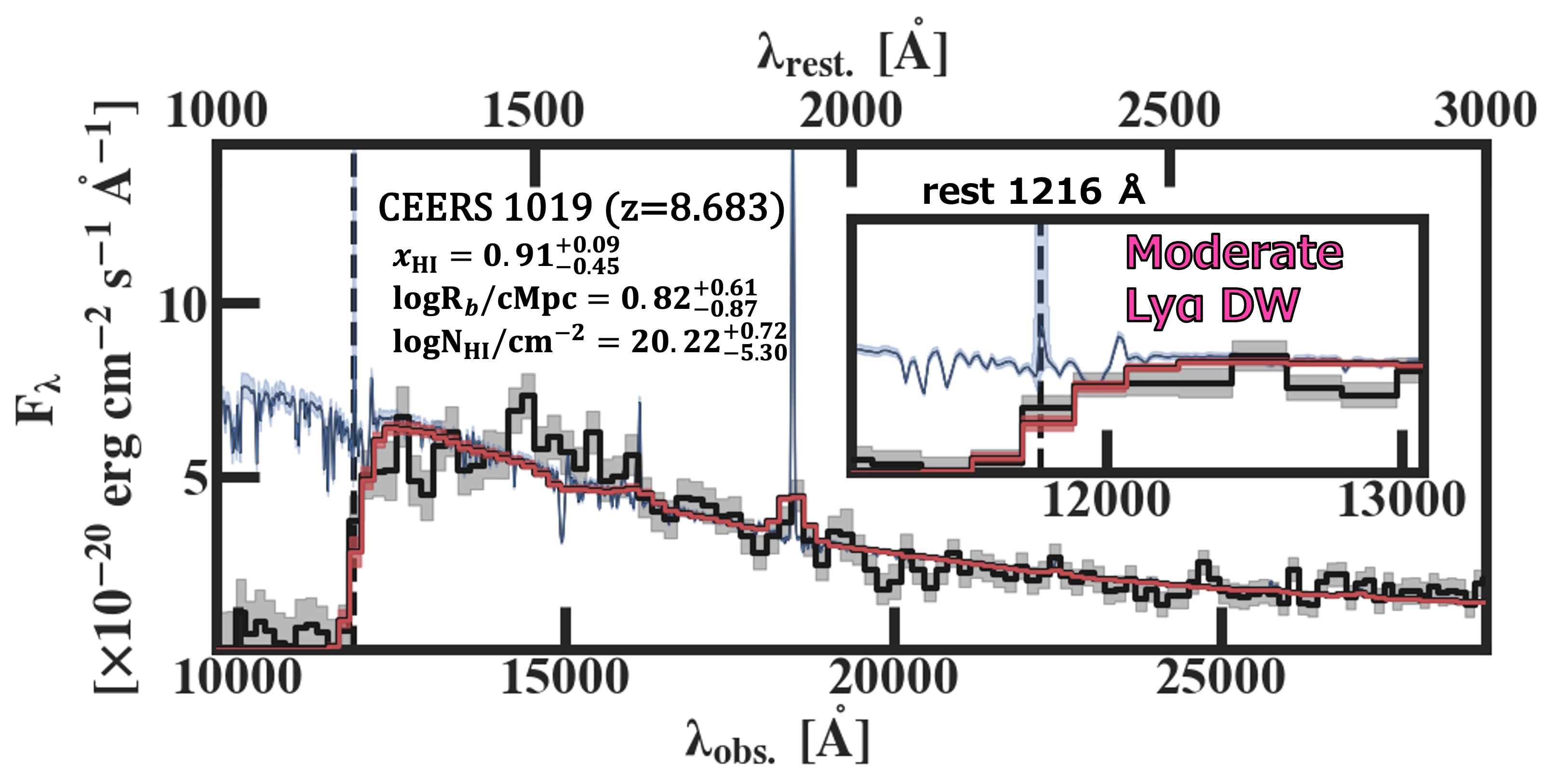}
\includegraphics[width=0.8\hsize,clip]{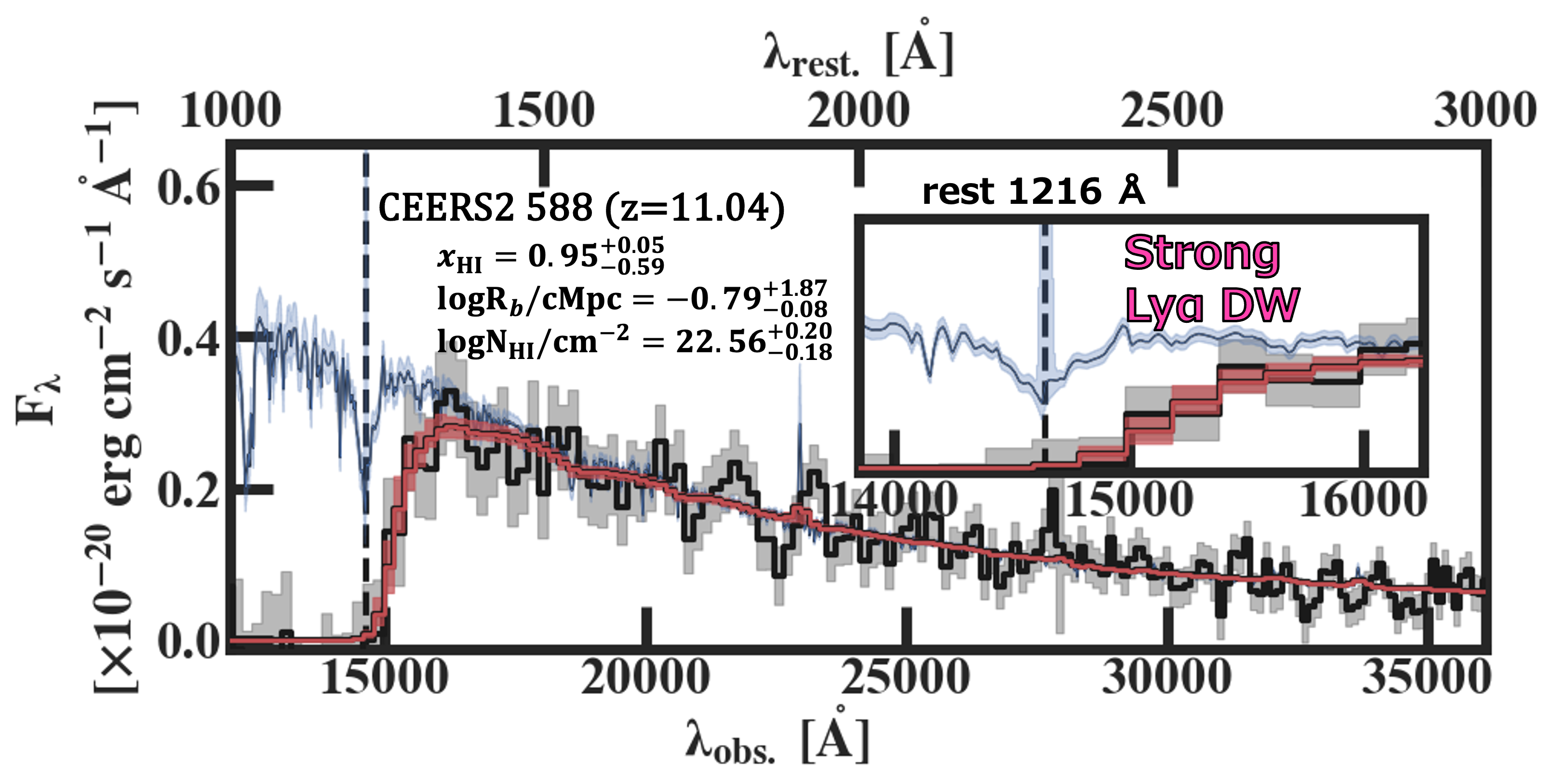}
\includegraphics[width=0.8\hsize,clip]{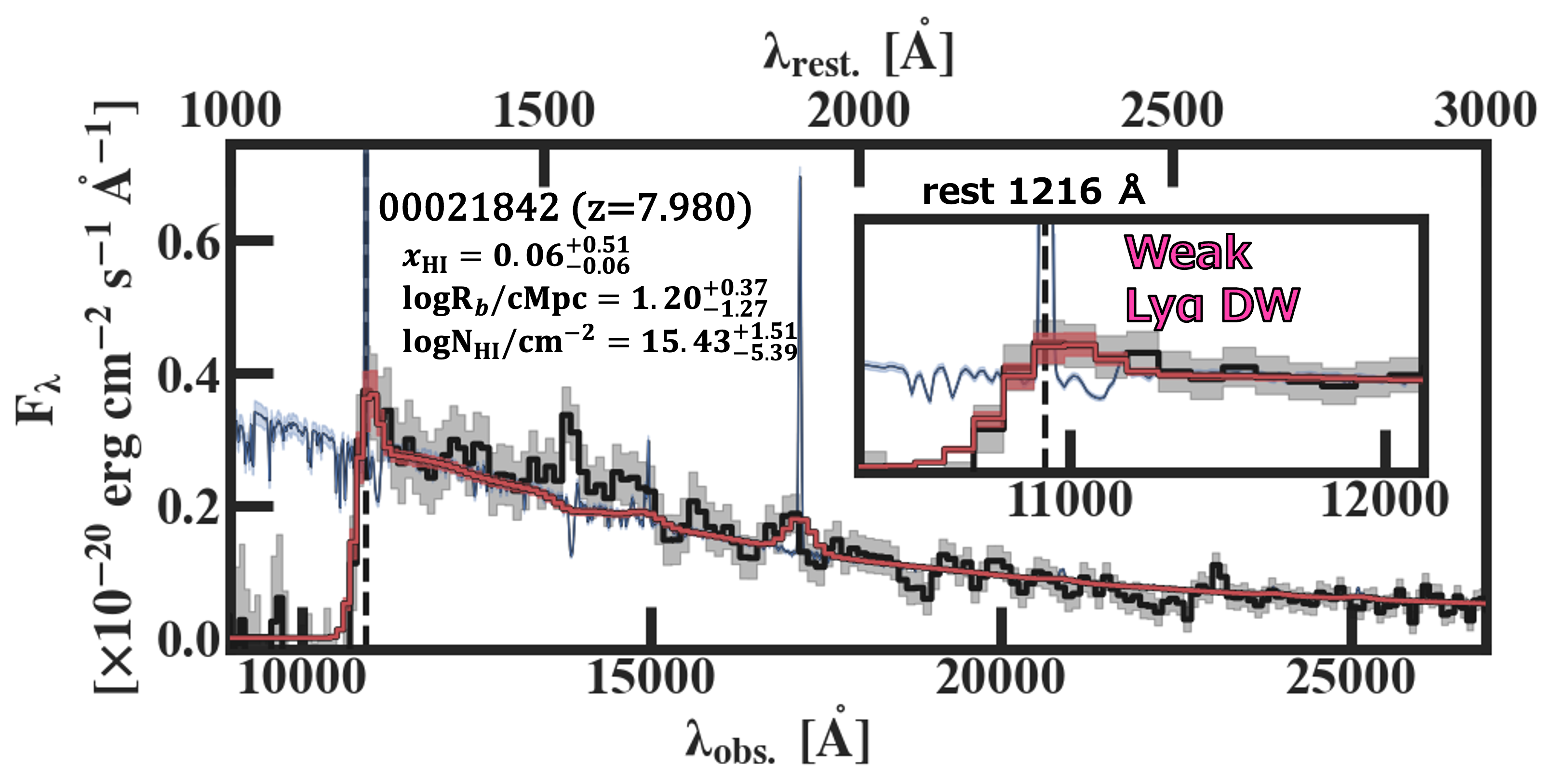}
\caption{\Add{Fitted spectrum for the case of moderate, strong and weak Ly$\alpha$ damping wing absorption feature seen in CEERS\_1019, CEERS2\_588, and 000012842 from the top to bottom panels, respectively. The black line and shade represent the observed spectrum and its error. The red (blue) line and shade represent the model spectrum and its uncertainty after (before) Ly$\alpha$ absorptions applied and smoothed with PRISM resolution. The black dotted line represent the rest-frame 1216 {\AA} at the systemic redshift. The inset panel shows zoom-in view of the spectrum around the rest-frame 1216 {\AA}. The best-fit values and uncertainties for $x_{\rm \HI}$, $\log R_b$, and $\log N_{\rm \HI}$ for each objects are also written inside each panel. The \Add{x-axes} at the top and bottom of each panel represent the observed and rest-frame wavelength, respectively.}}
\label{res:fit_spec}
\end{figure*}

\Add{In Figure \ref{res:fit_spec}, we also present examples of our spectrum fitting for individual objects. From the top to bottom panels, we show the fitting results for CEERS\_1019, CEERS2\_588, and 00021842, respectively. For CEERS\_1019, strong IGM absorption with $x_{\rm \HI}={0.91}^{+0.09}_{-0.45}$ and an ionized bubble size of $\log R_b/{\rm cMpc}={0.82}^{+0.61}_{-0.45}$ explains moderately softened break feature around the rest-frame 1216 {\AA} of the observed spectrum. For CEERS2\_588, in addition to strong IGM absorption (i.e., $x_{\rm \HI}\sim1$) and small ionized region (i.e., $\log R_b/{\rm cMpc}<0$), Ly$\alpha$ damping wing absorption by CGM with {\HI} column density of $\log N_{\rm \HI}/{\rm cm^{-2}}>22$ surrounding the galaxy explains strongly softened break feature around the rest-frame 1216 {\AA}. In contrast, the sharp and blue-shifted break feature at the rest-frame 1216 {\AA} for 00021842 can be explained with Ly$\alpha$ emission together with high Ly$\alpha$ transmission environment realized with small IGM absorption (i.e., $x_{\rm \HI}\sim0$) and a large ionized bubble (i.e., $\log R_b/{\rm cMpc}>1$).}

\begin{deluxetable}{cccc}
\tablecolumns{4}
\tabletypesize{\scriptsize}
\tablecaption{Best-fit Estimate for $x_{\rm \HI}$ and \Add{$\log R_b$}%
\label{table:bf}}
\tablehead{%
\colhead{ID} & %Object?
\colhead{$\langle z \rangle$} &
\colhead{$x_{\rm \HI}$} &
\colhead{$\log R_{b}$} \\
\colhead{} &
\colhead{} &
\colhead{} &
\colhead{[cMpc]} \\
\colhead{(1)} &
\colhead{(2)} &
\colhead{(3)} &
\colhead{(4)} 
}
\startdata
\#1 & $7.12^{+0.06}_{-0.08}$ & \Add{${0.53}^{+0.18}_{-0.47}$} & \Add{${1.67}^{+0.14}_{-0.16}$}\\
\#2 & $7.44^{+0.34}_{-0.24}$ & \Add{${0.65}^{+0.27}_{-0.34}$} & \Add{${1.53}^{+0.34}_{-1.48}$}\\
\#3 & $8.28^{+0.41}_{-0.44}$ & \Add{${0.91}^{+0.09}_{-0.22}$} & \Add{${0.70}^{+0.47}_{-1.04}$} \\
\#4 & \Add{$9.91^{+1.49}_{-1.15}$} & \Add{${0.92}^{+0.08}_{-0.10}$} & \Add{${-0.69}^{+0.89}_{-0.24}$}\\
\enddata
\tablecomments{
(1): ID for each subsample.
(2): Mean redshift and the lower/upper boundary of a subsample.
(3): \Add{Mode value and upper/lower limit of HPDI of $x_{\rm \HI}$ for the stacked marginalized PDFs in each redshift bin.} each redshift bin.
(4): \Add{Mode value and upper/lower limit of HPDI of $\log R_b$ for the stacked marginalized PDFs in each redshift bin.}
}
\end{deluxetable}

\section{Discussion} \label{disc}
\subsection{Ly$\alpha$ Absorption by CGM} \label{DLA}
\begin{figure*}[htbp]
\centering
\begin{center}
\includegraphics[width=\linewidth]{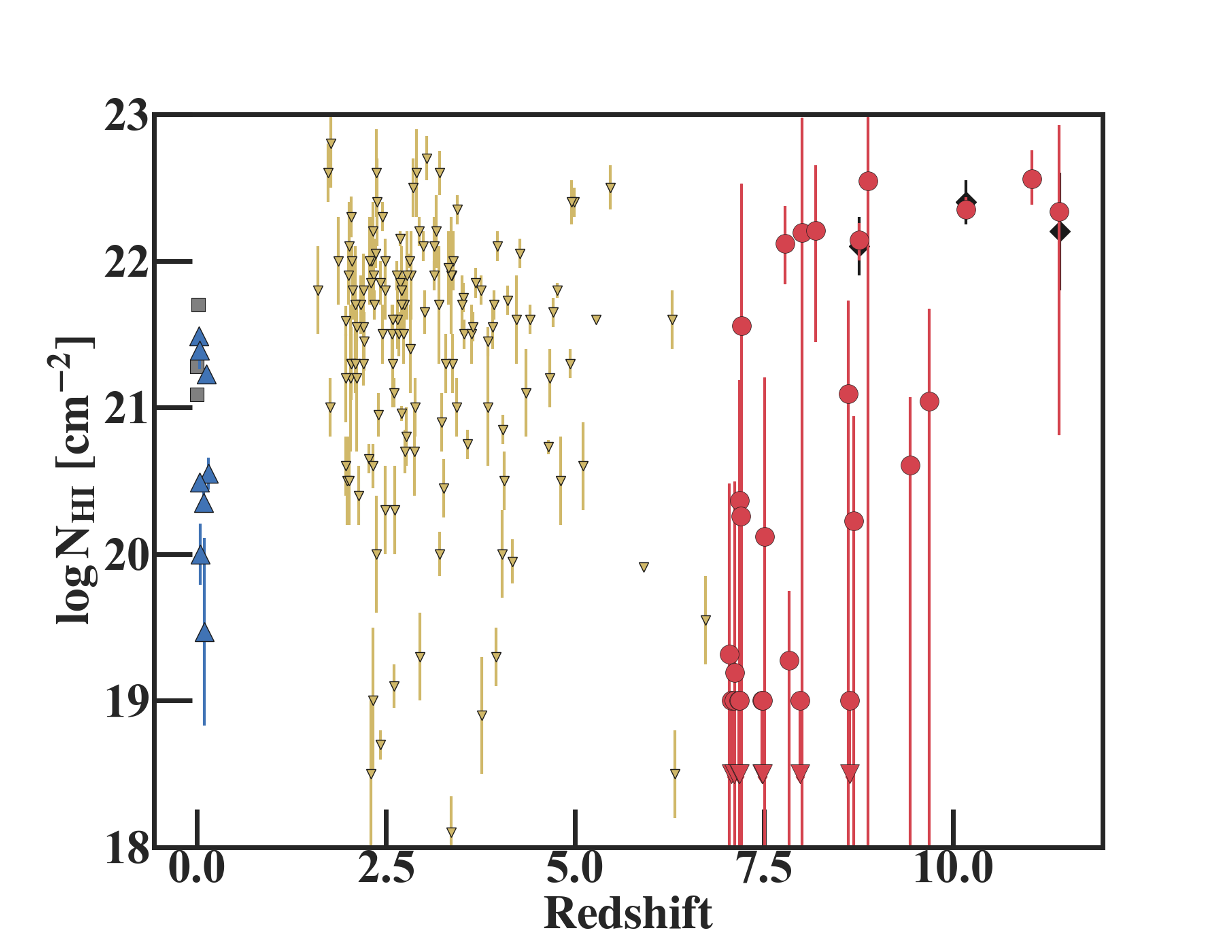}
\end{center}
\caption{\Add{Neutral hydrogen column density of Ly$\alpha$ absorption by \Add{{\HI} gas} at different redshifts. The red circles (bars) represent $\log N_{\rm \HI}$ measurements (uncertainties). The blue upward triangles, gray squares, and yellow downward triangles, and black diamonds represent the $\log N_{\rm \HI}$ measurements (uncertainties) of $z\sim0$ green pea galaxies \citep{2019ApJ...874...52M}, $z\sim0$ metal poor galaxies \citep{2020ApJ...892...19H}, gamma ray bursts \citep{2019MNRAS.483.5380T}, and $z>9$ Lyman break galaxies \citep{Heintz23}, respectively.}}
\label{logNHI}
\end{figure*}

\Add{In Figure \ref{logNHI}, we summarize the best-fit estimation of neutral hydrogen column density of {\HI} gas in CGM by redshift. In the same figure, we also present the measured neutral hydrogen column density of $z\sim0$ metal-poor galaxies \citep{2020ApJ...892...19H}, $z\sim0$ green pea galaxies \citep{2019ApJ...874...52M}, gamma ray bursts \citep{2019MNRAS.483.5380T}, and recent $z>8$ bright galaxies \citep{Heintz23}. Measured neutral hydrogen column density spans up to the range of damped Ly$\alpha$ absorbers (DLA; $N_{\rm \HI}>2\times10^{20}$). Moreover, neutral hydrogen column density reaches up to $\gtrsim10^{22}~{\rm cm}^{-2}$ for some objects, suggesting a existence of extreme reservoir of neutral hydrogen gas around some of the $z>7$ galaxies as suggested by \cite{Heintz23}. We compare the neutral hydrogen column density measurement of overlapping \Add{three} galaxy between our sample and \cite{Heintz23}'s and find that our measurements are consistent within errorbars with \cite{Heintz23}'s. This consistency further solidifies the validity of our analysis and the existence of high-z extreme DLA.}

\subsection{Escape Fraction} \label{fesc_disc}
\begin{figure}[htbp]
\centering
\begin{center}
\includegraphics[width=\linewidth]{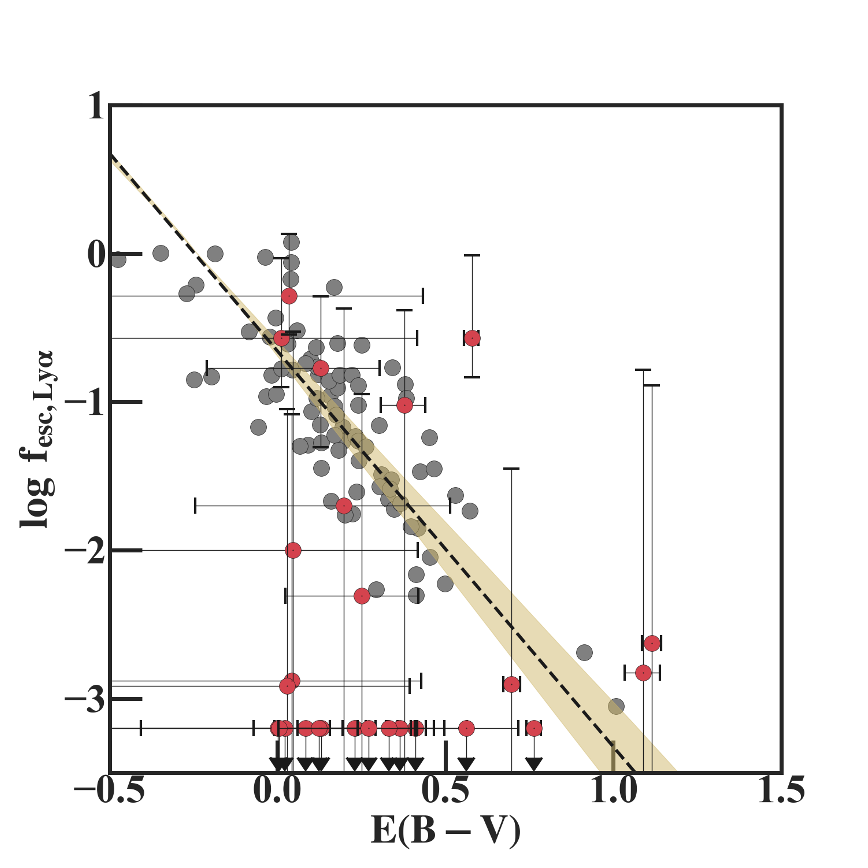}
\end{center}
\caption{\Add{Relation between $E(B-V)$ and Ly$\alpha$ escape fraction. The red circle and bars (arrows) represent best-fit and errors (upper limits) for our galaxy sample. The grey circle represent the measurement of $z\sim0$ galaxies presented in left panel of Figure 10 from \cite{2014A&A...561A..89A}. The black dotted line and yellow shades represent the best-fit relation and its errors between $E(B-V)$ and $f_{\rm esc, Ly\alpha}$ for $z\sim0$ galaxies as presented in \cite{2014A&A...561A..89A}.}}
\label{fesc}
\end{figure}
\Add{As the existence of extremely high neutral hydrogen column density CGM is suggested around some of the $z>7$ bright galaxies, we consider how much Ly$\alpha$ emission can escape to the IGM (i.e., Ly$\alpha$ escape fraction $f_{\rm esc, Ly\alpha}$). We estimate $f_{\rm esc, Ly\alpha}$ by comparing the Ly$\alpha$ emission flux after the absorption by CGM and the intrinsic Ly$\alpha$ emission flux. For the intrinsic Ly$\alpha$ emission flux, we use {\tt cloudy} prediction calculated by \cite{2017ApJ...840...44B} as {\tt Prospector} adopts. We note that we do not apply IGM absorption to calculate $f_{\rm esc, Ly\alpha}$. We also estimate $E(B-V)$ using sampled posterior PDFs of optical depth at 5500 {\AA} by dust attenuation. In Figure \ref{fesc}, we plot estimated $E(B-V)$ and $f_{\rm esc, Ly\alpha}$ and compare with local relation discussed in \cite{2014A&A...561A..89A}. As shown in Figure \ref{fesc}, best-fit value for our $f_{\rm esc, Ly\alpha}$ are below unity, which means that Ly$\alpha$ emission strength applied in our analysis does not violate prediction from photoionization models. Moreover, our estimates for $f_{\rm esc, Ly\alpha}$ are roughly consistent with the local relation between $E(B-V)$ and $f_{\rm esc, Ly\alpha}$, although some of our $f_{\rm esc, Ly\alpha}$ values are smaller than the local trend. Because emission line flux is smoothed in low-resolution PRISM spectra, $f_{\rm esc, Ly\alpha}$ cannot be well constrained for smaller values. Moreover, galaxy sample used to derive the relation in \cite{2014A&A...561A..89A} is partially selected based on the Ly$\alpha$ emission detection. In contrast, we construct our galaxy sample based on the brightness of continuum (i.e., UV magnitude), so it is natural to have our $f_{\rm esc, Ly\alpha}$ to lie on or below the local trend of \cite{2014A&A...561A..89A}. In any case, we have confirmed that our treatment of Ly$\alpha$ emission is physical and consistent with observational trend.}
\subsection{Cosmic Reionization History} \label{crh}
\begin{figure*}[htbp]
\centering
\begin{center}
\includegraphics[width=\linewidth]{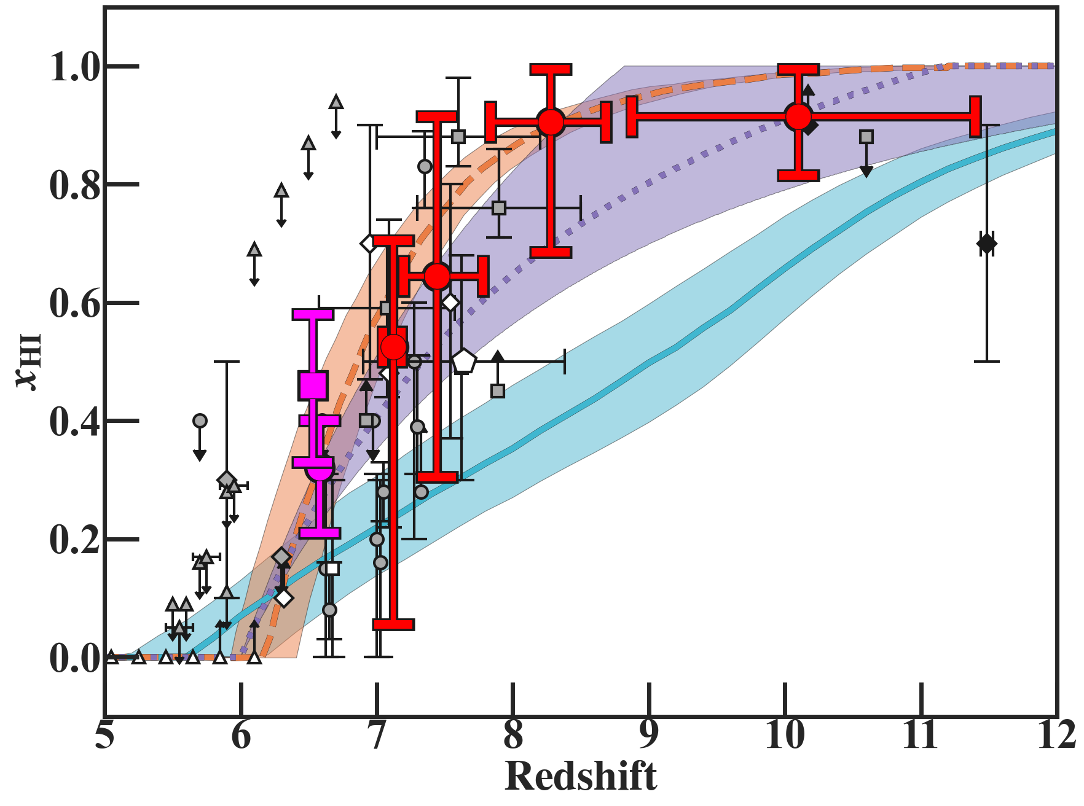}
\end{center}
\caption{Redshift evolution of $x_{\rm \HI}$. The red circle and bars represent the $x_{\rm \HI}$ estimates and the uncertainties from our work. The pink circle (square) and bars show the $x_{\rm \HI}$ estimate and uncertainty from $z=6-7$ Ly$\alpha$ luminosity function (LAE clustering analysis) with data obtained with Subaru/Hyper Suprime Cam \citep{U23_Thesis}. The orange dashed, blue solid, and purple dotted line and the shaded with corresponding colors are the model and its uncertainty for the redshift evolution of $x_{\rm \HI}$ predicted by \cite{N20}, \cite{F19}, and \cite{I18}, respectively. The grey circles and white squares represent the $x_{\rm \HI}$ estimation from Ly$\alpha$ luminosity function \citep{2010ApJ...723..869O,2014ApJ...797...16K,2018PASJ...70...55I,2018PASJ...70S..16K,2021ApJ...919..120M,2021ApJ...923..229G} and LAE clustering analysis \citep{2018PASJ...70S..13O}, respectively. The white and grey triangles represent the $x_{\rm \HI}$ estimation by QSOs Gunn Peterson troughs \citep{2006AJ....132..117F} and Ly$\alpha$+Ly$\beta$ dark fraction\Add{/gaps} \citep{2015MNRAS.446..566M,2023ApJ...942...59J,2022ApJ...932...76Z}, respectively. The grey and white diamonds represent the $x_{\rm \HI}$ estimation from Ly$\alpha$ damping wing absorption of QSOs \citep{2013MNRAS.428.3058S,2018ApJ...864..142D,2019MNRAS.484.5094G,2020ApJ...896...23W} and GRBs \citep{2006PASJ...58..485T,2014PASJ...66...63T}.
The black diamonds represent the $x_{\rm \HI}$ estimation from Ly$\alpha$ damping wing absorption profile of individual UV spectrum of galaxies \citep{CL23,H23}. 
The grey squares represent the $x_{\rm \HI}$ estimation from Ly$\alpha$ of Lyman-break galaxies \citep{2015MNRAS.446..566M,2019ApJ...878...12H,2019MNRAS.485.3947M,2020ApJ...904..144J,2020MNRAS.495.3602W,Br23,Mo23}. The white pentagon represents the $x_{\rm \HI}$ estimation from the cosmic microwave background observations \citep{2020A&A...641A...6P}.}
\label{history}
\end{figure*}
In Figure \ref{history}, we plot our ${x_{\rm \HI}}$ constraints and the uncertainties in red circle and bars, respectively. \Add{We also show literature values of $x_{\rm \HI}$ at different redshift including the constraint suggested by  the electron scattering from the cosmic microwave background observations \citep{2020A&A...641A...6P} and Ly$\alpha$ damping wing measurements of QSO spectra \citep{2013MNRAS.428.3058S,2018ApJ...864..142D,2019MNRAS.484.5094G,2020ApJ...896...23W}, GRB spectra \citep{2006PASJ...58..485T,2014PASJ...66...63T}, and recently obtained JWST galaxy spectra in others works \citep{CL23,H23}. Optical depth measurements using QSOs Gunn Peterson troughs around $z\sim5-6$ place the constraints on the end of the epoch of reionization \citep{2006AJ....132..117F}. We additionally show other $x_{\rm \HI}$ constraints from galaxy observations including the ones inferred from Ly$\alpha$ luminosity function \citep{2010ApJ...723..869O,2014ApJ...797...16K,2018PASJ...70...55I,2018PASJ...70S..16K,2021ApJ...919..120M,2021ApJ...923..229G}, Ly$\alpha$ emitter (LAE) clustering analysis \citep{2018PASJ...70S..13O}, Ly$\alpha$+Ly$\beta$ dark fraction\Add{/gaps} \citep{2015MNRAS.446..566M,2023ApJ...942...59J,2022ApJ...932...76Z}, and Ly$\alpha$ emission lines of Lyman-break galaxies \citep{2015MNRAS.446..566M,2019ApJ...878...12H,2019MNRAS.485.3947M,2020ApJ...904..144J,2020MNRAS.495.3602W,Br23,Mo23}.} For the $x_{\rm \HI}$ estimate from \cite{CL23}, we adopt the values shown in Figure 3 of the corresponding paper.\par

We overplot \Add{three distinct} evolutionary models by \cite{N20}, \cite{F19}, and \cite{I18} (hereafter, Naidu model, Finkelstein model, Ishigaki model, respectively) in orange dashed, blue solid, and purple dotted lines, respectively. We adopt Model II described in \cite{N20} as Naidu model. \Add{Note that the difference between redshift evolution of $x_{\rm \HI}$ in Model I and II in \cite{N20} is insignificant compared to relatively large errorbars for current $x_{\rm \HI}$ measurements.} These evolutionary models differ in the prescription for the escape fraction $f_{\rm esc}$ and the faint end slope $\alpha$ of the UV luminosity functions. Naidu model assumes that $f_{\rm esc}$ correlates with the star formation surface density $\Sigma_{\rm SFR}$ with a shallow faint end slope (i.e, $\alpha>-2$), resulting in a late and rapid cosmic reionization history driven by a small number of bright galaxies. Finkelstein model assumes that $f_{\rm esc}$ anti-correlate\Add{s} with halo mass $M_{\rm halo}$ and a steep faint end slope (i.e, $\alpha<-2$), resulting in an extended $x_{\rm \HI}$ evolution driven by many faint galaxies. Ishigaki model assumes the constant $f_{\rm esc}$ value at $\simeq 0.17$ with a steep UV faint end slope (i.e, $\alpha<-2$). Ishigaki model results in a moderately late cosmic reionization consistent with the one suggested by an electron scattering of the cosmic microwave background. We summarize the prescriptions for each model in Table \ref{table:model}. As shown in Figure \ref{history}, the \Add{best} estimate of our $x_{\rm \HI}$ \Add{consistent with all three evolutionary models and the constraints inferred from the electron scattering of CMB.}\par
In Figure \ref{history}, we also overplot the $x_{\rm \HI}$ values at $z\approx7$ estimated by \cite{U23_Thesis} from Ly$\alpha$ luminosity function and clustering properties of LAEs as pink circle and square, respectively. \cite{U23_Thesis} derive Ly$\alpha$ luminosity function and LAE spatial angular correlation function using the $z=6-7$ photometric LAE catalogs constructed by \cite{O21} from the Public Data Release 2 data of the Hyper Suprime-Cam Subaru Strategic Programs \citep{Ai19}. \cite{U23_Thesis} estimate $x_{\rm \HI}$ from Ly$\alpha$ luminosity function and angular correlation function of LAEs in a similar manner as \cite{2018PASJ...70S..16K}, \cite{2018PASJ...70S..13O}, \Add{and \cite{2021ApJ...923..229G}}. The $z=6-7$ LAE catalog of \cite{O21} is the largest in sample size with a total of 3561 galaxies, thus the $x_{\rm \HI}$ estimates from \cite{U23_Thesis} have advantages in terms of the reduced statistical uncertainty \Add{for estimated Ly$\alpha$ luminosity function and angular correlation functions}. \Add{For the $x_{\rm \HI}$ estimate from Ly$\alpha$ luminosity function, \cite{U23_Thesis} first compared the luminosity density at $z=5.7$ and $z=6.6$ to derive IGM transmission of Ly$\alpha$. Then, \cite{U23_Thesis} estimate $x_{\rm \HI}$ from the IGM transmission using the analytical relation between IGM transmission of Ly$\alpha$, $x_{\rm \HI}$, and the ionized bubble size from \cite{2014PASA...31...40D} and the semi-numerical relation between $x_{\rm \HI}$ and $R_b$ from \cite{FO05}. For the $x_{\rm \HI}$ estimate from LAE clustering analysis, \cite{U23_Thesis} compared the derived galaxy bias evolution from $z=5.7$ to $z=6.6$ to the semi-numerical prediction of galaxy bias at $z=6.6$ with different $x_{\rm \HI}$ by \cite{2006MNRAS.365.1012F}.} By comparing the $x_{\rm \HI}$ estimate at $z>7$ with the one at $z<7$ estimated by \cite{U23_Thesis}, we can \Add{confirm} smooth monotonic increase of $x_{\rm \HI}$ \Add{towards the higher redshift}. Other \Add{$x_{\rm \HI}$} constraints at $z>7$ from \Add{the} literature are mostly consistent with \Add{our measurements within error. To summarize, our $x_{\rm \HI}$ measurements at $z>7$ agree with various constraints, including the ones previously inferred from the luminosity function and clustering properties of LAEs at $z<7$, the electron scattering of CMB, and the evolution of UV luminosity function assuming a constant $f_{\rm esc}\sim0.2$.}
\begin{deluxetable}{cccc}
\tablecolumns{4}
\tabletypesize{\scriptsize}
\tablecaption{$x_{\rm \HI}$ Evolution Models%
\label{table:model}}
\tablehead{%
\colhead{Model} & %Object?
\colhead{Naidu} &
\colhead{Ishigaki} &
\colhead{Finkelstein} 
}
\startdata
$f_{\rm esc}$ & $\propto \Sigma_{\rm SFR}^{0.4} $ & 0.17 & $\propto M_{\rm halo}^{-1}$ \\
$\alpha$ &$>-2$ & $<-2$ & $<-2$\\
\enddata
\tablecomments{
Naidu, Ishigaki, and Finkelstein models corresponds to $x_{\rm \HI}$ evolution model presented in \cite{N20}, \cite{I18}, and \cite{F19}, respectively. We adopt Model II of the \cite{N20} as Naidu model.
}
\end{deluxetable}
\subsection{Bubble Size Evolution} \label{bubble}
\begin{figure*}[htbp]
\centering
\begin{center}
\includegraphics[width=\linewidth]{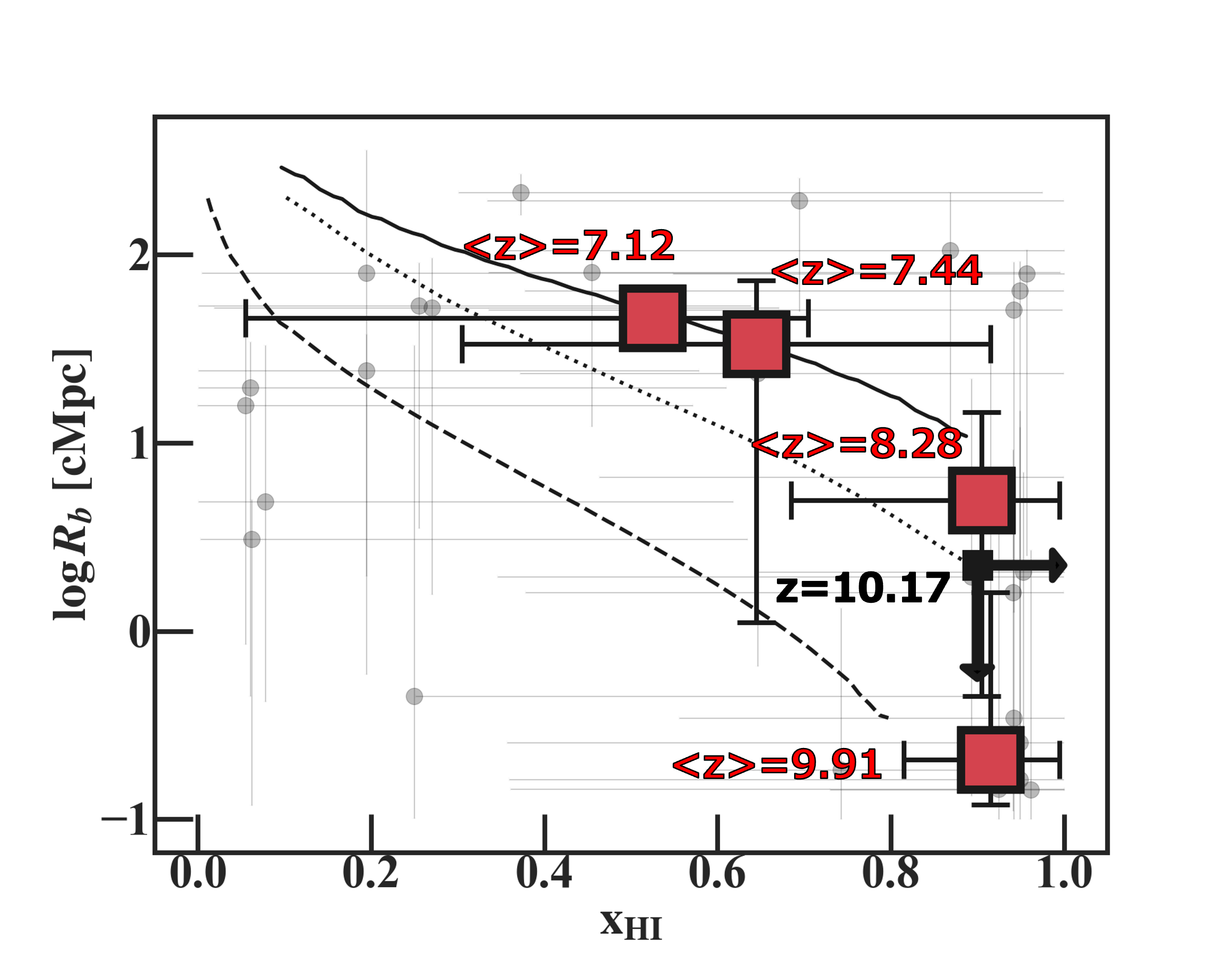}
\end{center}
\caption{$x_{\rm \HI}$ and $R_b$ estimates from different redshift bins. The red squares and bars represent our estimates and the errors, respectively, of ($x_{\rm \HI}$, $\log R_b$) at the redshift bin \#1, \#2, \#3, and \#4. Black square and arrows corresponds to 1$\sigma$ upper and lower limits of $x_{\rm \HI}$ and $R_b$, respectively, on MACS0647\_JD ($z=10.17$) by \cite{H23}. \Add{Faint grey circles and bars represent the best-fit value and its error for individual galaxies.} Solid and dashed lines correspond to the theoretical prediction by \cite{Lu23} on the characteristic ionized bubble radii at different $x_{\rm \HI}$ assuming rapid and gradual cosmic reionization considering overlapping ionized regions, respectively. The dotted line corresponds to the analytical relation between characteristic ionized bubble radii and $x_{\rm \HI}$ without considering overlapping ionized regions \citep{FO05}.}
\label{xHI_Rb}
\end{figure*}

Beside $x_{\rm \HI}$, the size of the ionized bubble is another tracer of cosmic reionization scenarios. Various studies suggest that in the case of inhomogeneous cosmic reionization driven by galaxies, the ionized region forms around the galaxies and their typical size correlates with $x_{\rm \HI}$ \citep[e.g.,][]{2003ApJ...596....1C,FO05,MF08a,MF08b,Lu23}. The size of ionized bubble at the EoR has been inferred from the Ly$\alpha$ emission line profiles and equivalent widths \citep[e.g.,][]{MG20,ES22,HS23,J23,Sax23}. We instead use Ly$\alpha$ damping wing profiles to investigate $R_b$ and its evolution with the redshift. We use the best-fit parameters presented in Section \ref{met} to see the relation between $R_b$ and $x_{\rm \HI}$. Figure \ref{xHI_Rb} shows the estimated $R_b$ values for each redshift bin from our analysis in Section \ref{met} together with the relation between $x_{\rm \HI}$ and $R_b$ predicted from analytical model by \cite{FO05} in black dotted lines. \Add{Together with our measurements, we show $x_{\rm \HI}$ and $R_b$ constrained via damping wing absorption analysis in \cite{H23}. We confirm that our constraint for the redshift bin of $\langle z \rangle=\Add{9.91}$ is consistent with $x_{\rm \HI}$ and $R_b$ measurements for $z=10.17$ object by \cite{H23}. In contrast, the recent bubble size measurement by \cite{2023arXiv230811609F} for $z=8.51$ object suggests a bubble size of $R_b=71.9\pm1.0$ cMpc. This bubble size value is about a dex larger than our constraint at the redshift bin of $\langle z \rangle=8.28$. As \cite{2023arXiv230811609F} discusses in their paper, this large offset of bubble size measurement could be physically attributed to specifically strong ionizing radiation by two possible AGN candidates with only a separation of 380 kpc within the same ionized bubble.} We also plot the $x_{\rm \HI}$ and $R_b$ relation for the rapid and gradual cosmic reionization history predicted from a semi-numerical simulations by \cite{Lu23} in solid and dashed lines, respectively. \cite{Lu23}'s \Add{predictions} take \Add{into} account of the overlap of the ionized regions \Add{formed around UV bright ($M_{\rm UV}<-20$) galaxies that are comparable to those for the majority of galaxies in our sample.} Incorporation of the overlaps between ionized yields systematically larger $R_b$ at each $x_{\rm \HI}$ compared with the ones predicted by \cite{FO05}, which do not consider the overlaps. As shown in Figure \ref{xHI_Rb}, our $R_b$ estimate at each $x_{\rm \HI}$ is \Add{up to few dex} larger than the one predicted by theoretical relations of \cite{FO05}, shown as the dotted line. However, our $R_b$ estimates are consistent with in errorbars with the ones predicted by \cite{Lu23} that consider the overlaps of ionized regions, confirming the importance of considering the overlaps to characterize the typical ionized bubble shapes during the EoR. Our inferred ionized bubble size is unexpected from the previous Ly$\alpha$ forest measurement. Ly$\alpha$ optical depth calculated based on our inferred $z\sim7-7.5$ ionized bubble radius is only found at $z\lesssim3$ by QSO Ly$\alpha$ forest measurements by \cite{2013MNRAS.430.2067B}. However, the Ly$\alpha$ optical depth measurement of \cite{2013MNRAS.430.2067B} does not incorporate contributions from high Ly$\alpha$ transmission at the proximity zone of the background QSO. Because our galaxies could be located at the overdensed region \Add{with a number of ionizing sources}, we must consider the proximity effect to have a fair comparison between our $z>7$ Ly$\alpha$ transmission estimate and those at $z<7$. \Add{When we compare our bubble size estimates to the previous measurement from QSO observations, our best-fit bubble radii estimates (i.e., $\sim10^2$ cMpc) at $z<8$ are even larger than the size of proximity zone around the brightest $z\sim6$ QSOs (i.e., $\lesssim10$ cMpc)\citep[e.g.,][]{2018ApJ...867...30E,2023arXiv230805800K}. Despite our bubble size estimates and the QSO proximity zone size does not have strong discrepancy because of the large errorbars for our measurement, one possibility to explain QSOs having smaller ionized region surrounding it than galaxy is that the QSO have shorter duty cycle than that of galaxies. Because gas needs to be irradiated by ionizing radiation long enough to form proximity zone, short QSO lifetime ($<10^5$ yr) has been proposed to explain small proximity observed at $z\sim6$ QSOs \citep{2018ApJ...864..142D}. The steady supply of ionizing photons through continuous star formation in galaxies could possibly enlarge the ionized bubble size more efficiently than young or short lived QSOs discovered at $z>6$.} Precise $R_b$ measurements in the future would be useful to distinguish different cosmic reionization scenarios \Add{(specifically on the formation process of ionized regions)} that yield different $x_{\rm \HI}$ and $R_b$ relations.

\subsection{Possible Systematics} \label{syst}
Although we consider the ionized bubbles surrounding the galaxies and \Add{{\HI} absorption by CGM} when we fit the spectra for the Ly$\alpha$ damping wing absorptions, there are still some possible systematics that we could consider for more accurate estimation. 
For example, the results from semi-numerical simulations suggest that our method assuming homogeneous IGM could overestimate the $x_{\rm \HI}$ values by $\sim30\%$ at the middle stage of cosmic reionization \citep{MF08a,MF08b}. \Add{\cite{2023arXiv230805800K} also suggest the simple analytical formulation using \cite{ME98} does miss out some of the physics related Ly$\alpha$ absorption that are seen in the simulation results (i.e., the contributions of residual {\HI} gas inside the ionized bubble to the Ly$\alpha$ optical depth). On the otherhand, the results from different numerical simulation with various box size yield different predictions on the evolution of key cosmic reionization parameters \citep[i.e., ][]{2023arXiv230805800K, Lu23}, indicating better interpretation of the Ly$\alpha$ damping wing absorption measurements requires further improvements in both simulation and observations.} \par
Another source of systematics is a biased sampling of galaxies \Add{in physical properties such as UV-luminosity.} \Add{Although we could use the simulation prediction suitable for the observed galaxies \citep[e.g., UV-bright galaxies residing at overdensed region;][]{Lu23} to infer cosmic reionization parameters,} we ideally need to know the optical depths of as many \Add{unbiased} sight lines as possible to precisely estimate the volume-averaged neutral fraction of the universe \Add{with a minimum dependency on the simulation models}. Moreover, we must investigate the evolution and diversity \Add{of physical properties} of the intrinsic UV spectrum of galaxies that are still not well studied above $z>7$ \Add{(e.g., UV-slope, ionizing photon production efficiency, etc.)} \Add{With increase in size of $z>7$ galaxy sample, we could construct composite galaxy spectra from galaxy subsamples with more homogeneous physical properties (i.e., $M_{\rm UV}$) as it has been done for $z<6$ galaxies \citep[e.g.,][]{Cu19,Cu20}. Analyzing a composite spectrum from a uniform galaxy sample has a potential for the estimation of average Ly$\alpha$ IGM absorption with high $S/N$ at the same time minimizing the impact of variations in intrinsic spectral shapes.}. In addition to the uncertainty in UV spectrum of galaxies, uncertainty in Ly$\alpha$ lines are also crucial in interpretation of the damping wing features seen in low-resolution PRISM spectra. \Add{Although most of the $z>7$ Ly$\alpha$ detected galaxies have the velocity offset similar to that for our model \citep[i.e., 200 km/s;][]{2023arXiv231206804N}, there are some reports of $z>7$ galaxies with large Ly$\alpha$ velocity offsets as large as $\sim500$ km/s \citep[e.g., ][]{Tan23}. Such a large Ly$\alpha$ velocity offset could result in higher Ly$\alpha$ transmission, which then mimic the spectra of galaxy surrounded by large ionized bubble in low-resolution data \citep[e.g.,][]{2023arXiv230805800K}. We conduct same analysis as described in Section \ref{dwfit} but with Ly$\alpha$ velocity offset fixed at 500 km/s, and confirmed that $x_{\rm \HI}$ and $\log R_{b}$ estimates for velocity offset at 200 km/s and 500 km/s are consistent within errorbars for all redshift bins.} The bubble size estimate for the two lower redshift bins (i.e., redshift bins with $\log R_b/{\rm cMpc}\sim2$ for fixed Ly$\alpha$ velocity offset at 200 km/s) with the velocity offset at 500 km/s change only by \Add{$\lesssim0.2$} in $\log R_b$ values, while errorbars remain as large as those for 200 km/s. This reflects the fact that we do not marginalize the intrinsic Ly$\alpha$ equivalent width over any other physical values, thus the effect of change in velocity offset is compensated by a change in the intrinsic Ly$\alpha$ emission strength. This demonstration suggests the importance of sophisticated Ly$\alpha$ emission modeling based on both observation and simulation for precise damping wing parameter measurements.\par
In terms of the systematics from data reduction, the damping wing analysis with low-resolution PRISM data introduces uncertainties in spectroscopic redshift and the degenerating components around Lyman break (e.g., Ly$\alpha$ emission, Ly$\alpha$ absorption by IGM/CGM). \Add{Additionally, as \cite{Jon23} shown, NIRSpec/PRISM spectra have systematics in the wavelength resolution depending on the compactness of the object. This could also be a potential source of the systematics in the redshift determination and damping wing absorption measurement using PRISM spectra.} \Add{The redshift uncertainty of $\Delta z\sim\pm0.01$ as suggested by \cite{Jon23} give a slight systematics of $\sim0.1$ for HPDI of $x_{\rm \HI}$ for the individual galaxy. This systematics is insignificant compared to the current large uncertainty of $x_{\rm \HI}$ ($\sim0.5$).}\par
\begin{figure}[htbp]
\centering
\begin{center}
\includegraphics[width=\linewidth]{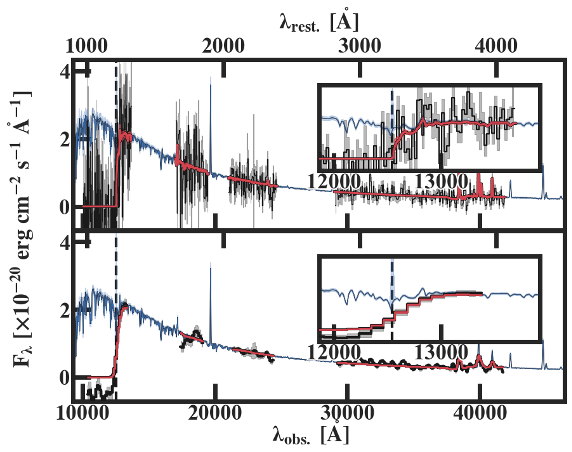}
\end{center}
\caption{\Add{Fitted spectrum of Gz9p3 in the high dispersion grating and PRISM data on the top and bottom panels, respectively. We show binned spectrum by 10 pixels for the visualization purpose.} The black line and shade represent the observed spectrum and its error. The red (blue) line and shade represent the model spectrum and its uncertainty after (before) Ly$\alpha$ absorptions applied and smoothed with instrumental broadening. The black dotted line represent the rest-frame 1216 {\AA} at the systemic redshift. The inset panel shows zoom-in view of the spectrum around the rest-frame 1216 {\AA}. The x-axes at the top and bottom of panels represent the observed and rest-frame wavelength, respectively.}
\label{Gz_spec}
\end{figure}
\begin{figure}[htbp]
\centering
\begin{center}
\includegraphics[width=\linewidth]{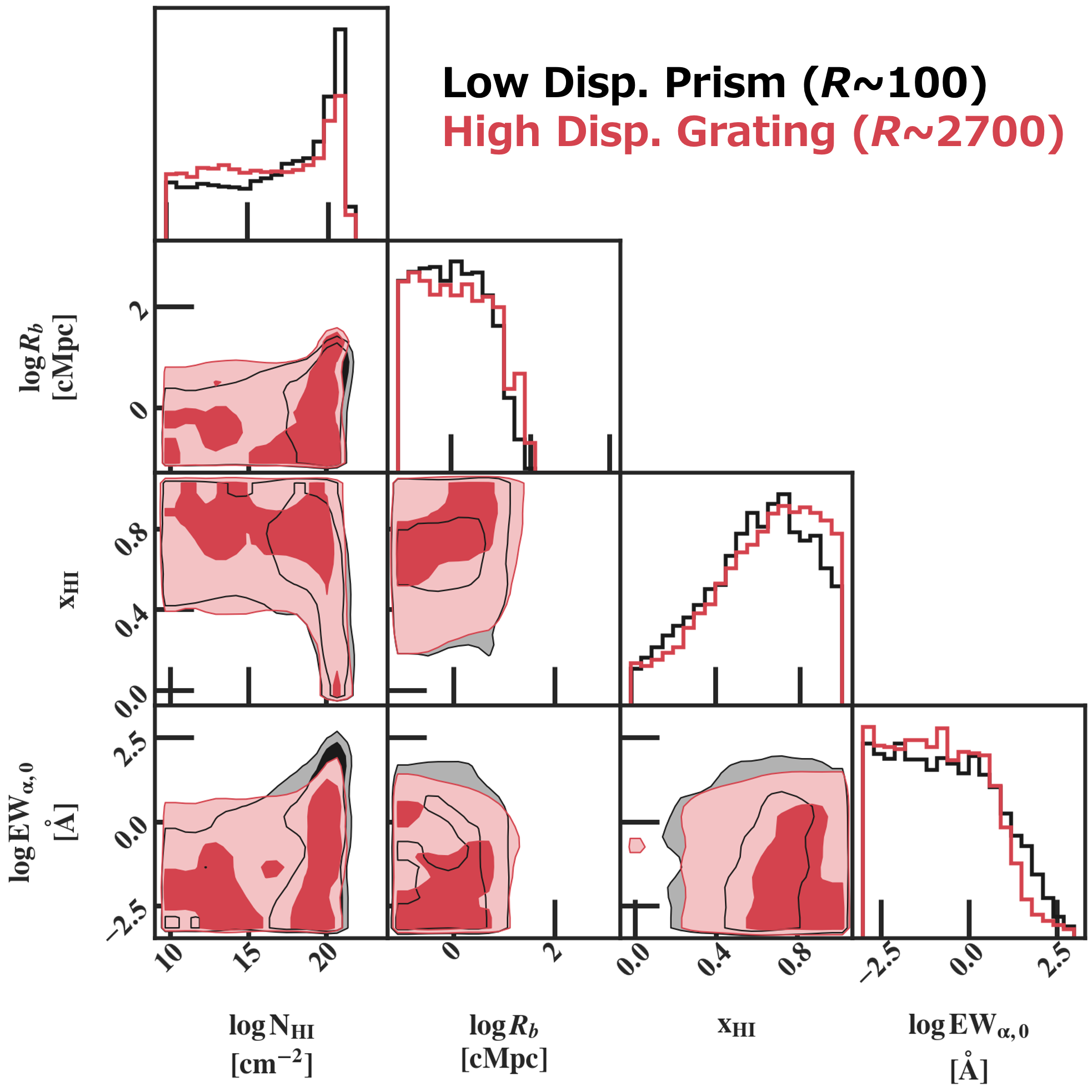}
\end{center}
\caption{\Add{The PDFs of Ly$\alpha$ emission and Ly$\alpha$ absorptions by {\HI} gas in the CGM/IGM for Gz9p3. The red and black colored mark corresponds to the result for high dispersion grating and PRISM data, respectively. Two shaded regions with different darkness in the bottom left panel correspond to the 1, and 2$\sigma$ regions of the sampled 2-D PDF from the darkest to the faintest, respectively. Top panel corresponds to the 1D marginalized PDF for each parameter.}}
\label{Gz_corner}
\end{figure}

\Add{Lastly, we investigate how the instrumental broadening in PRISM spectrum affect Ly$\alpha$ damping wing profile fitting using real observation data. Here, we consider a spectroscopically confirmed  galaxy at $z\sim9$ called Gz9p3 \citepe{Boy23b} as a test case. Gz9p3 is a suitable object to check how the instrumental broadening affect the parameter estimation for damping wing absorption, because its UV-continuum and Lyman break are detected in the high spectral resolution $(R\sim2700)$ data \citep{Boy23b, Ha23}.
JWST/NIRSpec data of Gz9p3 is obtained in the observation of GLASS \citep[ERS-1324; PI. T. Treu;][]{GLASS22}. Gz9p3's spectral data is taken in high dispersion grating filter pairs of F100LP/G140H, F170LP/G235H, and F290LP/G395H, with exposure time of 4.9 hours for each configuration. In this paper, we use the high dispersion grating data reduced by \cite{Ha23}. We adopt the spectroscopic redshift of $z=9.313$ determined using {\OII\W}3727 and {\NeIII\W}3869 \citep{Ha23}. We use 1D spectrum data in the wavelength range of the 1.00-1.37, 1.70-1.95, 2.10-2.47, and 2.90-4.19 $\mu$m. To generate the mock PRISM spectrum of Gz9p3, we forward model by conducting 1000 steps of Monte Carlo simulation. For each step of Monte Carlo simulation, we fluctuate the original high dispersion 1D spectrum according to the error spectrum and then convolve to the resolution of PRISM accordingly using the PRISM line spread function provided by STScI.
For both high dispersion grating and low dispersion prism data, we conduct the spectral fitting in the same manner as described in Section \ref{cfit}. 
We show observed and corresponding fitted model spectrum for high dispersion grating and PRISM data for in the top and bottom panel of Figure \ref{Gz_spec}, respectively. In Figure \ref{Gz_corner}, we show the posterior PDF of the parameter related to the Ly$\alpha$ damping wing absorption for Gz9p3. The results for high dispersion grating and PRISM data are shown in the black and red, respectively. As shown in Figure \ref{Gz_corner}, 1D/2D marginalized PDF for high dispersion grating and PRISM data are consistent within 68\% percentiles. Both high-dispersion grating and PRISM results suggests the existence of Ly$\alpha$ damping wing absorption feature suggested by \cite{Boy23b}. Note that even with high dispersion grating data, the precision of the parameter estimation does not increase. One possibility for the lack of improvement in the parameter estimation is that the signal to noise ratio of the high dispersion grating is still low. To summarize, we investigate how multiple systematics could affect our Ly$\alpha$ damping wing measurement. Obtaining more observational data and improvement on data reduction and simulating EoR galaxies are needed to further draw definitive conclusion on how cosmic reionization proceeds.
}

\section{Conclusion} \label{conc}
We present our constraints on the $x_{\rm \HI}$ and $R_b$ at $z=7-12$ via analysing Ly$\alpha$ damping wing absorption of \Add{27} galaxies. We gathered JWST/NIRSpec's PRISM data from multiple survey\Add{s} to create the galaxy sample. Ou\Add{r} sample consists of 2\Add{7} galaxies and the systemic redshifts of all galaxies have been confirmed with the emission lines including \Add{including {\OIII}5007, {\Hb}, and {\Hc}}. We then construct 4 composite spectra binned by the redshifts and confirm the \Add{softening break feature} near the rest-frame 1216 {\AA}, suggesting the strengthening of Ly$\alpha$ damping wing absorptions toward the high redshifts. We then fit the \Add{galaxy} spectra by the \Add{stellar model spectra} with Ly$\alpha$ damping wing absorptions that changes with $x_{\rm \HI}$ and $R_b$.
\begin{enumerate}
\item 
Stacked spectra at four redshift bins show the \Add{the transition from sharpened to softened break feature} around rest-frame 1216 {\AA} towards the high redshift, suggesting the increasing $x_{\rm \HI}$. We confirm that the median estimates for $x_{\rm \HI}$ \Add{($R_b$)} values for the stacked spectra at $\langle z \rangle=$\Add{7.12, 7.44, 8.28, and \Add{9.91}} show increasing (decreasing) trend toward the higher redshift bins.
\item 
The \Add{best-fit} estimates for $x_{\rm \HI}$ at $z>7$ by Ly$\alpha$ damping wing absorptions of galaxies' UV continua is consistent with moderately late reionization history that is consistent with the $x_{\rm \HI}$ constraints \Add{previously suggested} from CMB observations and the evolution of UV luminosity function with $f_{\rm esc}\sim0.2$. We note, however, that the errors of our $x_{\rm \HI}$ estimates are still large. We need to reduce statistical errors as well as possible systematics to narrow \Add{down the range of possible} cosmic reionization scenarios.
\item 
We compare estimated $x_{\rm \HI}$ and $R_b$ values with the theoretical predictions of the typical ionized bubble radii at different $x_{\rm \HI}$. Our $R_b$ estimates lie \Add{up to a few dex} above the cosmic average values estimated from analytic calculations by \cite{FO05}, while our $R_b$ estimates are comparable with the values for overlapping ionized bubbles around bright galaxies predicted by numerical simulations of \citep{Lu23}.
\end{enumerate}

\section*{Acknowledgements}\label{ack}
We thank Andrei Mesinger, Anne Hutter, Adi Zitrin, \Add{Akio Inoue, Hidenobu Yajima, Hayato Shimabukuro, Daichi Kashino, Satoshi Kikuta, Yoshinobu Fudamoto, Kazuyuki Omukai, Shintaro Yoshiura, Akinori Matsumoto, Minami Nakane}\Add{, Steven Finkelstein, Dan Stark, Xiaohui Fan, George Becker, and Yongda Zhu} for the valuable discussions on this work. 
\Add{We thank the editor of the journal and the anonymous referee for their detailed and thorough review of the paper that improved discussions of this work.}
This work is based on observations made with the NASA/ESA/CSA James Webb Space Telescope. 
The data were obtained from the Mikulski Archive for Space Telescopes at the Space Telescope Science Institute,
which is operated by the Association of Universities for Research in Astronomy, Inc., under NASA contract NAS 5-03127 for JWST.
These observations are associated
with programs ERS-1345 (CEERS), GO-1433, DDT-2750, \Add{GTO-1210 (JADES)}, and \Add{ERS-1324 (GLASS)}.
The authors acknowledge the CEERS, GO-1433, DDT-2750, \Add{JADES, and GLASS} teams led by Steven L. Finkelstein, Dan Coe, Pablo Arrabal Haro, \Add{Nora Lützgendorf}, and \Add{Tomasso Treu}, respectively,
for developing their observing programs with a zero-exclusive-access period. We thank the JADES team for publicly releasing reduced spectra and catalog from the JADES survey. We thank Fergus Cullen and the team for publicly releasing the composite UV spectra from VANDELS survey.
This publication is based upon work supported by the World Premier International Research Center Initiative (WPI Initiative), MEXT, Japan, and KAKENHI
(20H00180, 21J20785, 21K13953, 21H04467, 23KJ0646, 23KJ0589, 22K21349) through Japan Society for the Promotion of Science. \Add{This work was supported by JSPS Core-to-Core Program (grant number: JPJSCCA20210003).} This work was supported by the joint research program of the Institute for Cosmic Ray Research (ICRR), University of Tokyo. This research was supported by FoPM, WINGS Program, the University of Tokyo.

\bibliography{cite.bib}{}
\bibliographystyle{aasjournal}

\appendix

\section{Posterior Probability Distribution Functions Including Nuisance Parameters}

\begin{figure*}
\centering
\begin{minipage}{0.49\hsize}
\begin{center}
\includegraphics[width=0.99\hsize,clip]{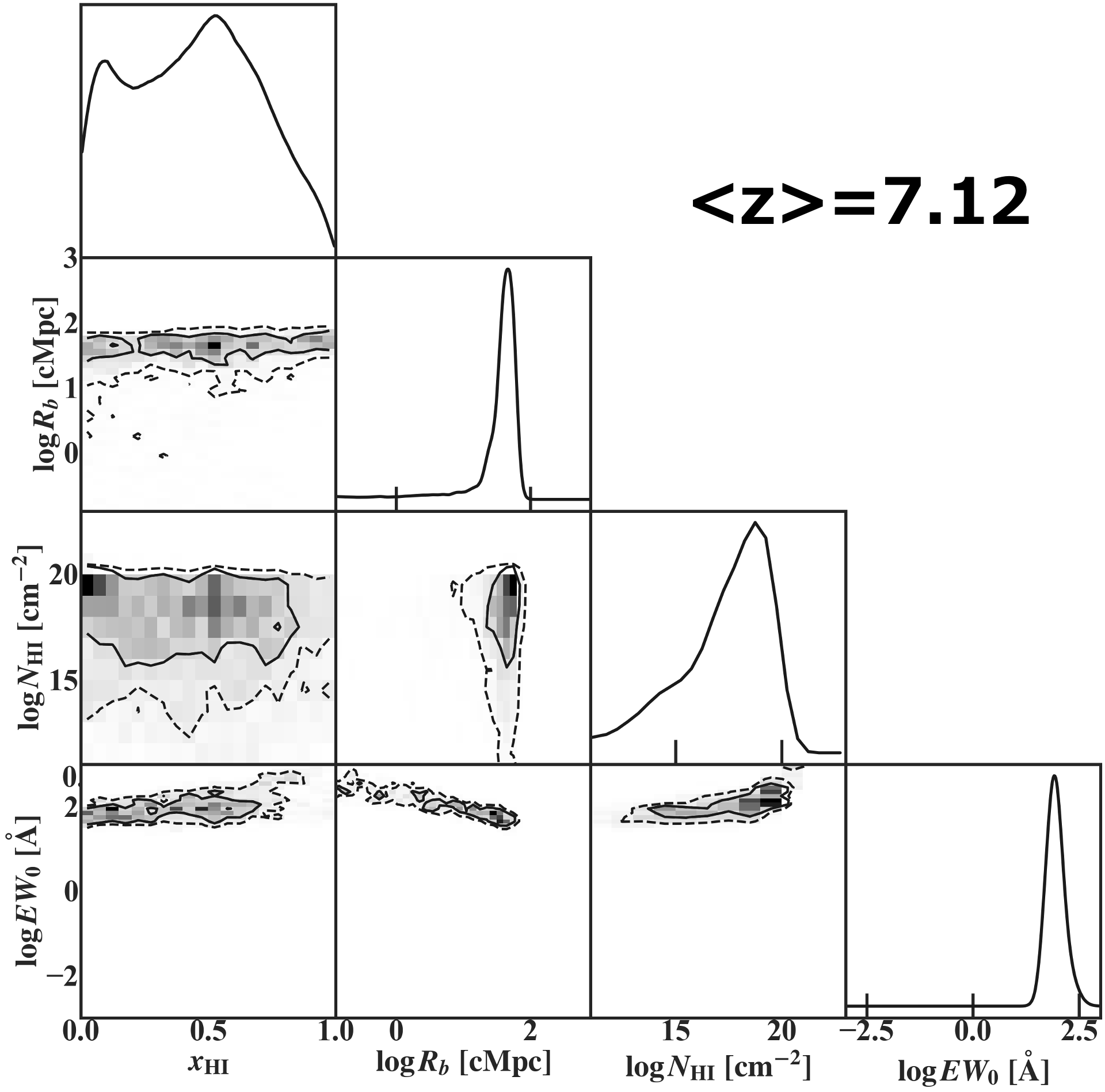}
\end{center}
\end{minipage}
\begin{minipage}{0.49\hsize}
\begin{center}
\includegraphics[width=0.99\hsize,clip]{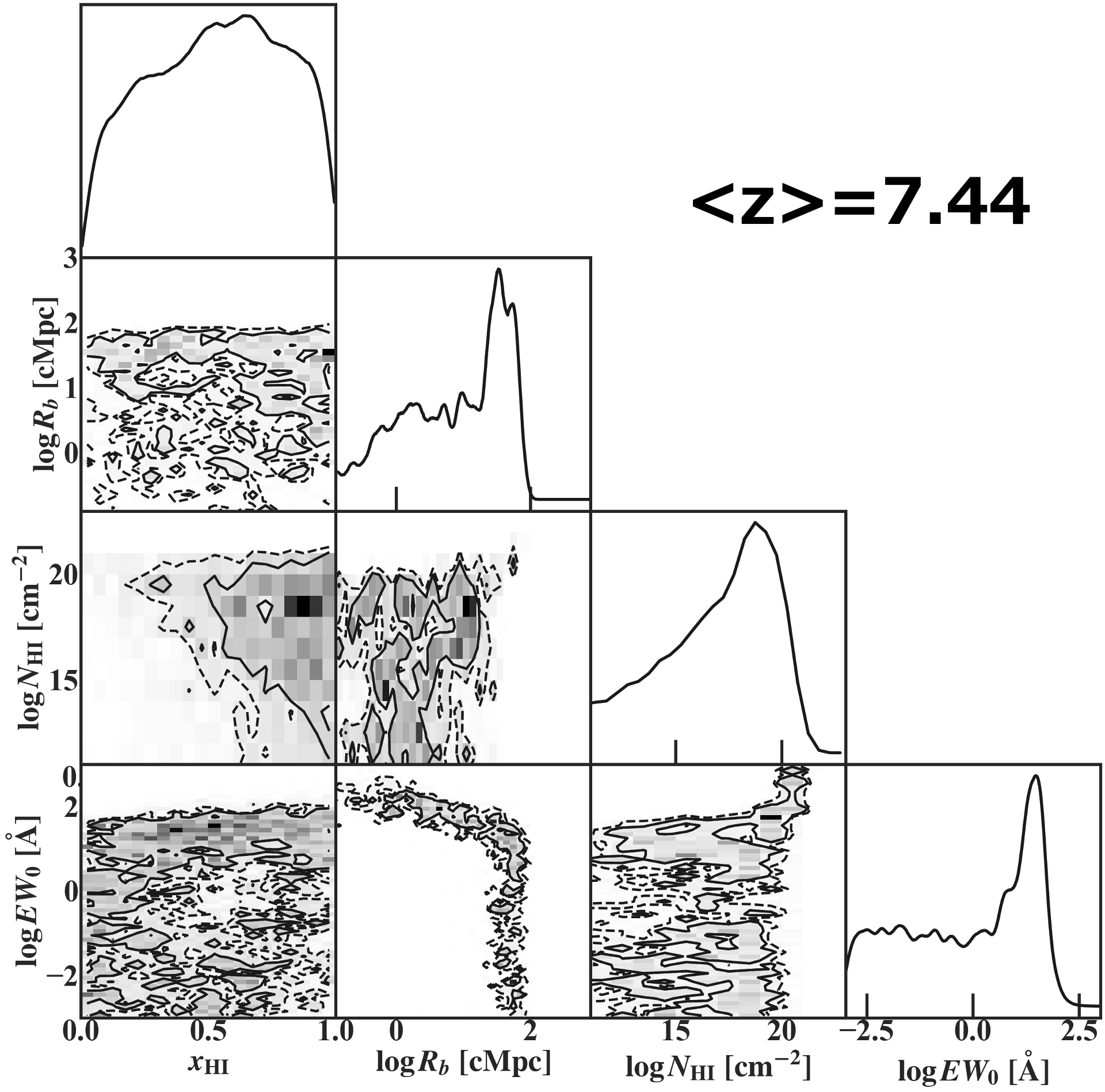}
\end{center}
\end{minipage}

\par
\vspace{0.5cm}

\begin{minipage}{0.49\hsize}
\begin{center}
\includegraphics[width=0.99\hsize,clip]{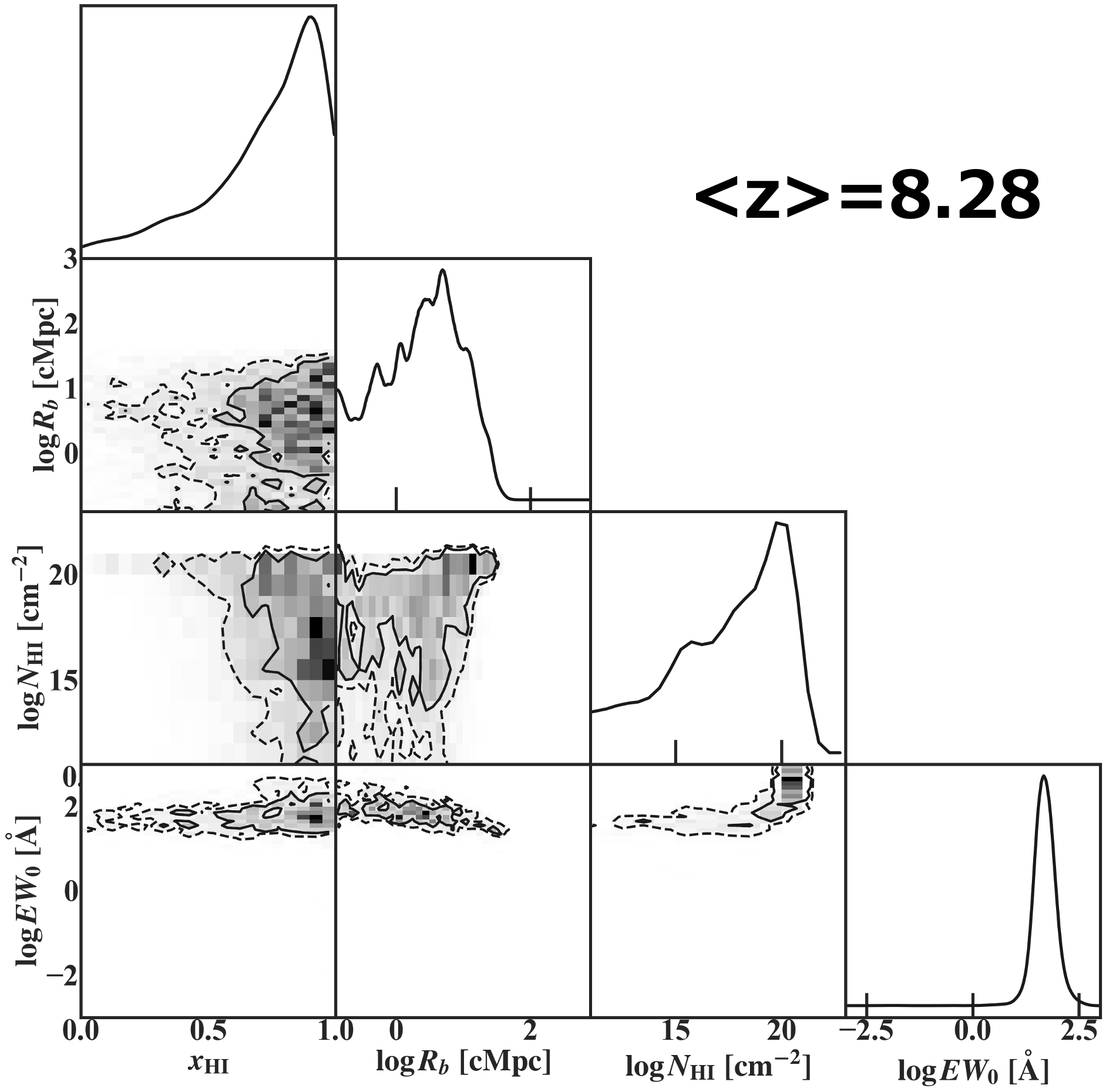}
\end{center}
\end{minipage}
\begin{minipage}{0.49\hsize}
\begin{center}
\includegraphics[width=0.99\hsize,clip]{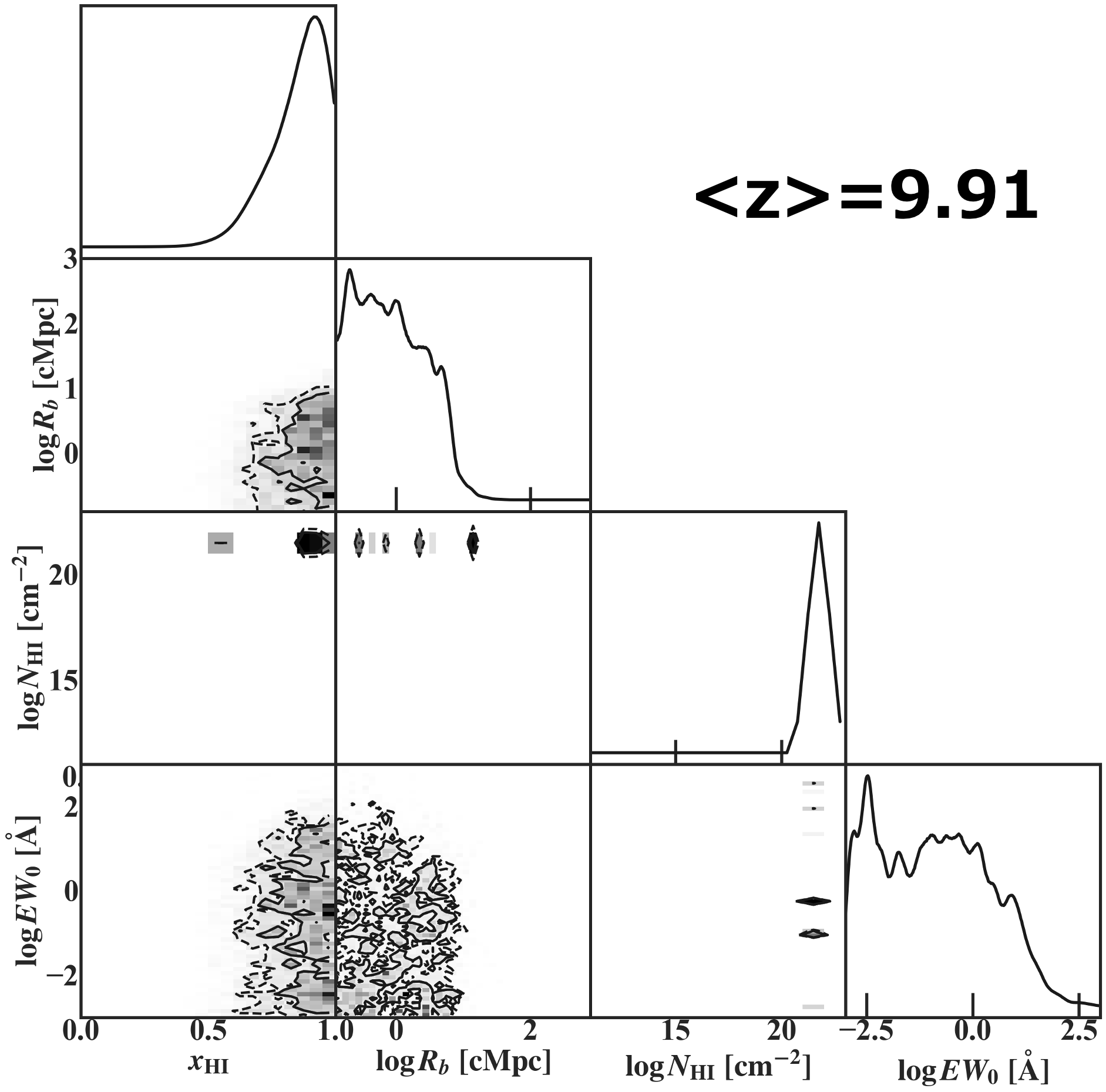}
\end{center}
\end{minipage}

\caption{\Add{The posterior probability distribution function (PDF) of Ly$\alpha$ emission and Ly$\alpha$ absorptions by {\HI} gas in the CGM/IGM for each subsample. The results for \#1  to \#4 bin are shown in order of top left to right bottom. The darkness in the 2D-marginalized PDFs represent the probability density. The solid and dotted contours represent the 68 and 90-th percentiles, respectively. Top panel corresponds to the 1D marginalized PDF for each parameter. The presented PDFs are smoothed by gaussian kernel.}}
\label{res:fit_res_full}
\end{figure*}

\Add{To briefly highlight the importance of considering the intrinsic galaxy's property on precise inference of globally defined parameters (e.g., $x_{\rm \HI}$, $\log R_b$), we show the posterior PDFs of global IGM parameters together with the parameters related to the intrinsic Ly$\alpha$ strength and the host galaxy's $\HI$ absorption strength in Figure \ref{res:fit_res_full}. We derive the posterior PDFs for $\log EW_{\alpha,0}$ and $\log N_{\rm \HI}$ in the same manner as for $x_{\rm \HI}$ and $\log R_b$. We note, however, that $\log EW_{\alpha,0}$ and $\log N_{\rm \HI}$ are nuisance parameters and unique to the individual galaxies. Therefore, we only refer binned posterior PDFs for $\log EW_{\alpha,0}$ and $\log N_{\rm \HI}$ in terms of how the nuisance parameters could affect the inference of $x_{\rm \HI}$ and $\log R_b$. As an example, the marginalized 2-D posterior PDF for $\log R_b$ and $\log EW_{\alpha,0}$ for $<z>=7.44$ shows clear correlation between the two parameters. This correlation can be explained by how strong Ly$\alpha$ emission could mimic the high transmission (i.e., large ionized bubble), which leads to the degeneracy between the two parameters \citep[e.g.,][]{2023arXiv230805800K} and the large uncertainty.}

\end{document}